\documentclass[a4paper,11pt]{article}
\pdfoutput=1
\usepackage{jheppub}

\usepackage[T1]{fontenc} 

\usepackage{amssymb,amsmath,bm,bbm,natbib}
\usepackage{color}
\usepackage{slashed}
\usepackage{graphics}
\usepackage{graphicx}
\usepackage[utf8]{inputenc}
\usepackage[caption=false]{subfig}
\usepackage{url}
\usepackage{dsfont}
\usepackage{float} 
\usepackage{cancel}
\usepackage{xcolor,colortbl}
\usepackage{rotating}
\usepackage[export]{adjustbox}
\usepackage{subfig}

\newcommand{\eq}[1]{Eq.~\eqref{#1}}
 
\newcommand{\Br}{\text{Br}}
\newcommand{\Ft}{\mathcal{F}t}

\newcommand{\hc}{\text{h.c}}

\def\XXint#1#2#3{{\setbox0=\hbox{$#1{#2#3}{\int}$}
		\vcenter{\hbox{$#2#3$}}\kern-0.5\wd0}}

\newcommand{\e}{\ensuremath{\mathrm{e}}}

\definecolor{Gray}{gray}{0.7}

\renewcommand{\arraystretch}{1.3}

\preprint{CERN-TH-2020-128, PSI-PR-20-11,  UZ-TH 26/20}
	\title{\boldmath Global Electroweak Fit and Vector-Like Leptons in Light of the Cabibbo Angle Anomaly}

\author[a,b,c]{Andreas Crivellin}

\author[b,c]{Fiona Kirk}
		
\author[b,c]{Claudio Andrea Manzari}

\author[b,c]{Marc Montull}

\affiliation[a]{CERN Theory Division, CH--1211 Geneva 23, Switzerland}
\affiliation[b]{Physik-Institut, Universit\"at Z\"urich, Winterthurerstrasse 190, CH--8057 Z\"urich, Switzerland}
\affiliation[c]{Paul Scherrer Institut, CH--5232 Villigen PSI, Switzerland}
	
\emailAdd{andreas.crivellin@cern.ch}
\emailAdd{fiona.kirk@psi.ch}
\emailAdd{claudioandrea.manzari@physik.uzh.ch}
\emailAdd{marc.montull@psi.ch}
\abstract{
	The "Cabibbo Angle Anomaly" (CAA) originates from the disagreement between the CKM elements $V_{ud}$ and $V_{us}$ extracted from superallowed beta and kaon decays, respectively, once compared via CKM unitarity. It points towards new physics with a significance of up to $4\,\sigma$, depending on the theoretical input used, and can be explained through modified $W$ couplings to leptons. In this context, vector-like leptons (VLLs) are prime candidates for a corresponding UV completion since they can affect $W\ell\nu$ couplings at tree-level, such that this modification can have the dominant phenomenological impact. In order to consistently assess agreement data, a global fit is necessary which we perform for gauge-invariant dimension-6 operators and all patterns obtained for the six possible representations (under the SM gauge group) of VLLs. We find that even in the lepton flavour universal case, including the measurements of the CKM elements $V_{us}$ and $V_{ud}$ into the electroweak fit has a relevant impact, shifting the best fit point significantly. 
	Concerning the VLLs we discuss the bounds from charged lepton flavour violating processes and observe that a single representation cannot describe experimental data significantly better than the SM hypothesis. However, allowing for several representations of VLLs at the same time, we find that the simple scenario in which $N$ couples to electrons via the Higgs and $\Sigma_1$ couples to muons not only explains the CAA but also improves the rest of the electroweak fit in such a way that its best fit point is preferred by more than $4\,\sigma$ with respect to the SM.
}

\begin{document} 
\maketitle
\flushbottom

\newpage

\section{Introduction}\label{Introduction}

While new physics (NP) coupling to quarks or gluons is strongly constrained by direct LHC searches (see e.g. Refs.~\cite{Butler:2017afk,Masetti:2018btj} for an overview), there is much more parameter space left for models with new particles posessing only electroweak (EW) interactions. 
In this context, vector-like leptons (VLLs), which are heavy fermions that are neutral under QCD and can mix with SM leptons via Higgs interactions, are very interesting. VLLs are predicted in many SM extensions, such as Grand Unified Theories~\cite{Hewett:1988xc,Langacker:1980js,delAguila:1982fs}, composite models or models with extra dimensions~\cite{Antoniadis:1990ew,ArkaniHamed:1998kx,Csaki:2004ay,ArkaniHamed:2001nc,ArkaniHamed:2002qy,Perelstein:2005ka,delAguila:2010vg,Carmona:2013cq} and, last but not least, are involved in the type I~\cite{Minkowski:1977sc,Lee:1977tib} and type III~\cite{Foot:1988aq} seesaw mechanisms. In fact, as expected,
LEP~\cite{Achard:2001qw} and LHC~\cite{Aad:2019kiz,Sirunyan:2019ofn}\footnote{For a recent dedicated theoretical analysis of VLLs at colliders, see e.g.~\cite{Chala:2020odv,Das:2020gnt, Das:2020uer,deJesus:2020upp}.} searches allow for VLLs with masses far below the TeV scale. Therefore, it is well possible that VLLs are the lightest states within a NP model superseding the SM, thus providing the dominant NP effects in the EW sector of the SM. Note that even by simply adding by hand any VLL to the SM one obtains a consistent UV complete (renormalizable and anomaly free) extension of it, that can thus be studied on its own.
\smallskip

Since VLLs can couple to SM leptons and the Higgs, they mix with the former after EW symmetry breaking~\cite{Langacker:1988ur}. This mixing modifies the couplings of the SM leptons to EW gauge bosons ($W$ and $Z$), which are tightly constrained by LEP measurements~\cite{Schael:2013ita,ALEPH:2005ab}. In particular, any modification of the $W\ell\nu$ coupling is always accompanied by an effect in the $Z\ell\ell$ and/or $Z\nu\nu$ couplings. Furthermore, $W\mu\nu$ and $W e\nu$ couplings affect
the extraction of the Fermi constant $G_F$ from muon decay. Therefore, their impact on different observables is clearly correlated and in order to consistently study them, it is necessary to perform a global fit to all the EW observables. This was done previously in Ref.~\cite{delAguila:2008pw} for all the VLL representations and in Refs.~\cite{Antusch:2014woa,deGouvea:2015euy,Fernandez-Martinez:2016lgt,Chrzaszcz:2019inj,Crivellin:2020lzu} for the VLLs corresponding to the type I or type III seesaw. However, since the publication of Ref.~\cite{delAguila:2008pw} the experimental situation has changed significantly. In particular, the Higgs mass is now known~\cite{Aaboud:2018wps,Sirunyan:2017exp} and the top~\cite{TevatronElectroweakWorkingGroup:2016lid,Khachatryan:2015hba,Sirunyan:2018gqx} and $W$~\cite{Aaltonen:2012bp,D0:2013jba,Aaboud:2017svj} mass measurements have become much more precise.
\smallskip

Furthermore, recently the ``Cabibbo Angle Anomaly'' (CAA) has emerged with a significance of up to $4\,\sigma$~\cite{Belfatto:2019swo,Grossman:2019bzp,Coutinho:2019aiy,Crivellin:2020lzu,Endo:2020tkb}. 
This anomaly is due to the disagreement between the CKM element $V_{us}$ extracted from kaon and tau decays, and the one determined from beta decays, in particular super-allowed beta decays (using CKM unitarity). 
One can consider this discrepancy to be a sign of (apparent) CKM unitarity violation~\cite{Belfatto:2019swo,Cheung:2020vqm}. 
However, a sizable violation of CKM unitarity is in general difficult to generate due to the strong bounds from flavour-changing neutral currents, such as kaon mixing (see e.g.\ Ref.~\cite{Bobeth:2016llm}). 
Alternatively, one can consider the CAA as a sign of lepton flavour universality (LFU) violation (LFUV)~\cite{Coutinho:2019aiy,Crivellin:2020lzu,Capdevila:2020rrl,Endo:2020tkb}. In fact, flavour dependent modified neutrino couplings to the $W$ and $Z$ gauge bosons provide a very good fit to the data~\cite{Coutinho:2019aiy} and this view seems to be a natural since experiments have accumulated intriguing hints for the violation of LFU within recent years. 
In particular, the measurements of the ratios $R(D^{(*)})$~\cite{Lees:2012xj,Aaij:2017deq,Abdesselam:2019dgh} and $R(K^{(*)})$~\cite{Aaij:2017vbb,Aaij:2019wad} deviate from the SM expectation of LFU by more than $3\,\sigma$~\cite{Amhis:2019ckw,Murgui:2019czp,Shi:2019gxi,Blanke:2019qrx,Alok:2019uqc} and $4\,\sigma$~\cite{Alguero:2019ptt,Aebischer:2019mlg,Ciuchini:2019usw,Arbey:2019duh}, respectively. 
The anomalous magnetic moments $(g-2)_\ell$ of the charged leptons are also a measure of LFU violation as they vanish in the massless limit. 
Here, the long-standing discrepancy of about $3.7\,\sigma$~\cite{Bennett:2006fi,Aoyama:2020ynm} in the anomalous magnetic moment of the muon, $(g-2)_\mu$,~\footnote{Recently, the BMWc released a lattice calculation of hadronic vacuum polarisation in $(g-2)_\mu$ whose results would bring theory and experiment of $(g-2)_\mu$ into agreement. However, this result disagrees with $e^+e^-$ to hadron data~\cite{Davier:2017zfy,Keshavarzi:2018mgv,Davier:2019can,Keshavarzi:2019abf,Colangelo:2018mtw,Ananthanarayan:2018nyx} and would increase the tension in the EW fit~\cite{Crivellin:2020zul,Keshavarzi:2020bfy} as hadronic vacuum polarisations contribute to the running of $\alpha$, which, at the scale $M_Z$, is a crucial input for the EW fit. We checked that modified gauge boson couplings to leptons are not capable of reducing this tension significantly and we therefore use the result from $e^+e^-$ to hadrons.} and the more recently emerging deviation of $2.5\,\sigma$ in the anomalous magnetic moment of the electron, $(g-2)_e$, interestingly, with the opposite sign, could have a common origin~\cite{Davoudiasl:2018fbb,Crivellin:2018qmi}. In fact, it has been shown in Refs.~\cite{Czarnecki:2001pv,Kannike:2011ng,Dermisek:2013gta,Freitas:2014pua,Aboubrahim:2016xuz,Kowalska:2017iqv,Raby:2017igl,Megias:2017dzd,Calibbi:2018rzv,Crivellin:2018qmi,Arnan:2019uhr} that $(g-2)_\mu$ of the muon can be explained by VLLs, and in Refs.~\cite{Gripaios:2015gra,Arnan:2016cpy,Raby:2017igl,Arnan:2019uhr,Kawamura:2019rth} VLLs are involved in the explanation of $b\to s\ell^+\ell^-$ via loop effects.
\smallskip 

We take these developments as a motivation to perform an updated global EW fit~\cite{Haller:2018nnx,deBlas:2016ojx} to the modified EW gauge boson couplings to leptons. In particular, we want to assess the impact of including the $V_{us}$ and  $V_{ud}$ measurements in the fit and see if an explanation of the CAA is possible. We will do this first in a model independent way by performing a fit to the dimension-6 operators which (directly) change the lepton's gauge boson couplings. 
Then we perform a fit to all six representations of VLLs. Here, also contributions to flavour changing decays of charged leptons (such as $\mu\to e\gamma$, $\mu\to3e$, the analogous tau decays, and $\mu\to e$ conversion) can arise, which we calculate and analyse as well.  
\smallskip

This article is structured as follows: in the next section we will establish our setup, before calculating the contributions to the relevant observables and discussing the experimental situation in Sec.~\ref{observables}. In Sec.~\ref{analysis} we will perform our global fit, first in a model independent fashion including dimension-6 operators, 
and after for each of the six representations of VLLs separately. Finally, we conclude in Sec.~\ref{Conclusions}.
\section{Setup}
Let us establish our setup by first considering the effective dimension-6 operators (in the Warsaw basis) 
that generate modified $W\ell\nu$, $Z\nu\nu$ and $Z\ell\ell$ couplings after EW symmetry breaking. We will then turn to the six possible representations of VLLs under the SM gauge group and perform the matching on the effective operators.
\subsection{EFT}
Disregarding magnetic operators whose effect vanishes at zero momentum transfer and which can only be generated at the loop level, there are three operators (not counting flavour indices) in the $SU(3)_c\times SU(2)_L\times U(1)_Y$-invariant SM EFT which (directly) modify the couplings of neutrinos and charged leptons to the EW gauge bosons~\cite{Buchmuller:1985jz,Grzadkowski:2010es}.
\begin{equation}
\mathcal{L} = \mathcal{L}_{SM} + \dfrac{1}{\Lambda^2}\left( C_{\phi \ell}^{\left( 1 \right) ij} Q_{\phi \ell}^{\left( 1 \right) ij}  + C_{\phi \ell}^{\left( 3\right) ij} Q_{\phi \ell }^{\left( 3 \right) ij} + C_{\phi e}^{ij} Q_{\phi e }^{ij}\right)\,,
\label{Lagrangian}
\end{equation}
with
\begin{equation}
\begin{aligned}
Q_{\phi \ell }^{\left( 1 \right)ij} &= {\phi ^\dag }i{{\mathord{\buildrel{\lower3pt\hbox{$\scriptscriptstyle\leftrightarrow$}} 
			\over D} }_\mu }\phi \, {{\bar \ell_L}^i}{\gamma ^\mu }{\ell_L^j}\,,\\
Q_{\phi \ell }^{\left( 3 \right)ij} &= {\phi ^\dag }i\mathord{\buildrel{\lower3pt\hbox{$\scriptscriptstyle\leftrightarrow$}} 
	\over D} _\mu ^I\phi  \, {{\bar \ell_L}^i}{\tau ^I}{\gamma ^\mu }{\ell_L^j}\,,\\
Q_{\phi e}^{ij} &= {\phi ^\dag }i{{\mathord{\buildrel{\lower3pt\hbox{$\scriptscriptstyle\leftrightarrow$}} 
			\over D} }_\mu }\phi \, {{\bar e_R}^i}{\gamma ^\mu }{e_R^j}\,,
\end{aligned}
\label{eq:ops}
\end{equation}
where 
\begin{equation}
D_{\mu}=\partial_{\mu}+ig_2W_{\mu}^a \tau^a+ig_1B_{\mu}Y\,.\label{CovD}
\end{equation}
Here $i$ and $j$ are flavour indices and the Wilson coefficients $C$ are dimensionless. The operators defined in \eq{eq:ops} result in the following  modifications of the $Z$ and $W$ boson couplings to leptons after EW symmetry breaking
\begin{equation}
\mathcal{L}_{W,Z}^{\ell,\nu}=\bigg({{\bar \ell }_f}\Gamma _{fi}^{\ell\nu}{\gamma ^\mu }{P_L}{\nu _i}\,{W_\mu } + h.c.\bigg)+ \left[ {{{\bar \ell }_f}{\gamma ^\mu }\left( {\Gamma _{fi}^{\ell L}{P_L} + \Gamma _{fi}^{\ell R}{P_R}} \right){\ell _i} + {{\bar \nu }_f}\Gamma _{fi}^\nu {\gamma ^\mu }{P_L}{\nu _i}} \right]{Z_\mu }\,,
\label{definitionZll}
\end{equation}
with
\begin{equation}
\begin{aligned}
\Gamma _{fi}^{\ell L} &= \frac{{{g_2}}}{{2{c_W}}}\left[ {\left( {1 - 2s_W^2} \right){\delta _{fi}} + \frac{{v^2}}{{\Lambda^2}}\left( {C_{\phi \ell }^{\left( 1 \right)fi} + C_{\phi \ell }^{\left( 3 \right)fi}} \right)} \right]\,,\\
\Gamma _{fi}^{\ell R} &= \frac{{{g_2}}}{{2{c_W}}}\left[ { - 2s_W^2{\delta _{fi}} + \frac{{v^2}}{{\Lambda^2}}C_{\phi e}^{fi}} \right]\,,\\
\Gamma _{fi}^\nu  &=  - \frac{{{g_2}}}{{2{c_W}}}\left( {{\delta _{fi}} + \frac{{v^2}}{{\Lambda^2}}\left( {C_{\phi \ell }^{\left( 3 \right)fi} - C_{\phi \ell }^{\left( 1 \right)fi}} \right)} \right)\,,\\
\Gamma _{fi}^{\ell\nu} &=  - \frac{{{g_2}}}{{\sqrt 2 }}\left( {{\delta _{fi}} + \frac{{v^2}}{{\Lambda^2}}C_{\phi \ell }^{\left( 3 \right)fi}} \right)\,,
\end{aligned}
\label{Gammas}
\end{equation}
Here we used the convention $v/\sqrt{2}\approx 174\,$GeV. Eqs.~\eqref{definitionZll} and \eqref{Gammas} agree with Ref.~\cite{Dedes:2017zog}. The terms proportional to the Kronecker delta correspond to the (unmodified) SM couplings. 
\medskip

\subsection{Vector-Like Leptons}

Moving beyond the model independent approach of the last subsection, we now consider models with VLLs. As mentioned in the introduction, these particles modify the $Z$ and $W$ couplings to leptons already at tree-level and can therefore give dominant effects in the corresponding observables entering the global EW fit, in particular in the determination of $V_{us}$ and $V_{ud}$, related to the CAA. 
\smallskip

We define VLLs as fermions whose left and right-handed components have the same representations of $SU(2)_L\times U(1)_Y$, are singlets under QCD and can couple to the SM Higgs and SM leptons via Yukawa-like couplings. The possible representations under the SM gauge group are given in Table~\ref{VLLrep}. Since these fermions are vectorial, they can have bare mass terms (already before EW symmetry breaking) and interact with SM gauge bosons via the covariant derivative which was defined in \eq{CovD}.\footnote{In the case $\psi$ equals $N$ or $\Sigma_0$, which are Majorana fermions, i.e. $N_R=N_L^c$ or $\Sigma_{0,R}=\Sigma_{0,L}^c$, \eq{Lquad} should be defined with a factor ${1}/{2}$ to ensure a canonical normalisation.}
\begin{align}
\mathcal{L}^{\rm VLL} = \sum_\psi \, i \, \bar{\psi} \gamma_{\mu}D^{\mu}\,\psi -  M_{\psi}\,\bar{\psi}\psi\,,
\label{Lquad}
\end{align}
with $\psi=N,E,\Delta_1,\Delta_3,\Sigma_1,\Sigma_3$. The interactions of the VLLs with the SM leptons are given by
\begin{align}
-\mathcal{L}_{NP}^{\rm int} =&\, \lambda_N^i\, \bar{\ell}_i\,\tilde{\phi}\, N + \lambda_E^i\, \bar{\ell}_i\,\phi\, E + \lambda_{\Delta_1}^i\, \bar{\Delta}_1\,\phi\, e_i +  \label{Lint} \\
&\lambda_{\Delta_3}^i\, \bar{\Delta}_3\,\tilde{\phi}\, e_i + \lambda_{\Sigma_0}^i\, \tilde{\phi}^{\dagger}\,\bar{\Sigma}_0^I\,\tau^I\,  \ell_i + \lambda_{\Sigma_1}^i\, \phi^{\dagger}\,\bar{\Sigma}_1^I\,\tau^I\,  \ell_i +{\rm h.c.}\,,\nonumber
\end{align}
where $i$ is a flavour index and $\tau^I=\sigma^I/2$ are the generators of $SU(2)_L$. Here we neglected interactions of two different VLL representations with the Higgs\footnote{These couplings which would induce mixing among the VLLs are in general not important with respect to the modified $Z$ and $W$ couplings studied in this article, as they only give rise to dim-8 effects here. However, they can have important phenomenological consequences in magnetic dipole operators, allowing for an explanation of the $(g-2)_{\mu,e}$ via chiral enhancement~\cite{Czarnecki:2001pv,Kannike:2011ng,Dermisek:2013gta,Freitas:2014pua,Aboubrahim:2016xuz,Kowalska:2017iqv,Raby:2017igl,Megias:2017dzd,Calibbi:2018rzv,Crivellin:2018qmi,Arnan:2019uhr}.}. Our conventions for the VLL-triplets after EW symmetry breaking are
\begin{align}
\Sigma_0 = \frac{1}{2}\begin{pmatrix} \Sigma_0^0 & \sqrt{2}\Sigma_0^+ \\ \sqrt{2}\Sigma_0^- & -\Sigma_0^0  \end{pmatrix}, \quad
\Sigma_1 = \frac{1}{2}\begin{pmatrix} \Sigma_1^- & \sqrt{2}\Sigma_1^0 \\ \sqrt{2}\Sigma_1^{--} & -\Sigma_1^-  \end{pmatrix}\,,
\end{align}
where the superscript labels the electric charge.
\smallskip

\begin{table}[t!]
	\centering
	\begin{tabular}{l | c c c  } & $SU(3)$& {$SU(2)_L$}&$U(1)_Y$\\
		\hline
		$\ell$ &1 & 2 & -1/2 \\
		e &1 & 1 & -1 \\
		$\phi$ &1 & 2 & 1/2 \\
		\hline
		N &1 & 1 & 0 \\
		E & 1& 1 & -1 \\
		$\Delta_1= (\Delta_1^0, \Delta_1^-)$ & 1 & 2 & -1/2\\
		$\Delta_3 = (\Delta_3^-, \Delta_3^{--})$ & 1 & 2 &-3/2 \\
		$\Sigma_0 = (\Sigma_0^+, \Sigma_0^0, \Sigma_0^- )$ & 1 & 3 & 0 \\
		$\Sigma_1= (\Sigma_1^0, \Sigma_1^-, \Sigma_1^{--} )$& 1 & 3 & -1
	\end{tabular}	\caption{Representations of the SM leptons ($\ell,e$), the SM Higgs Doublet ($\phi$) and the VLLs under the SM gauge group.
}\label{VLLrep}
\end{table}

\begin{figure}[t]
	\centering
	\includegraphics[width=0.42\textwidth]{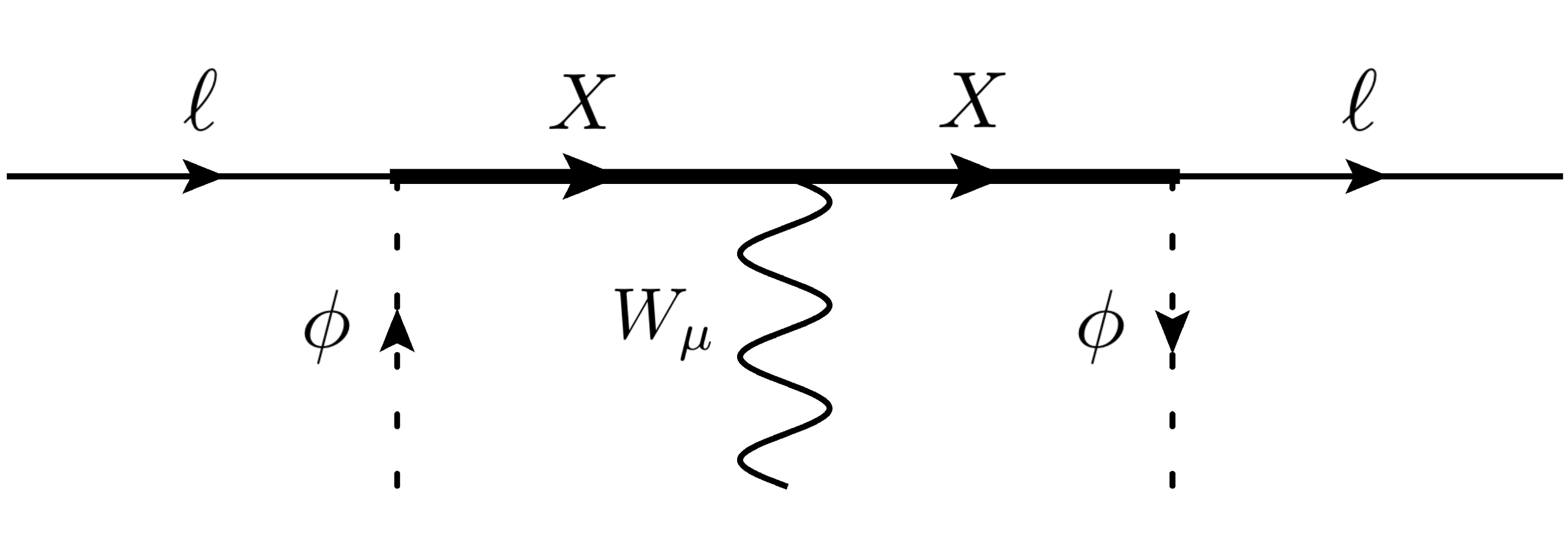} \\
	\includegraphics[width=0.42\textwidth]{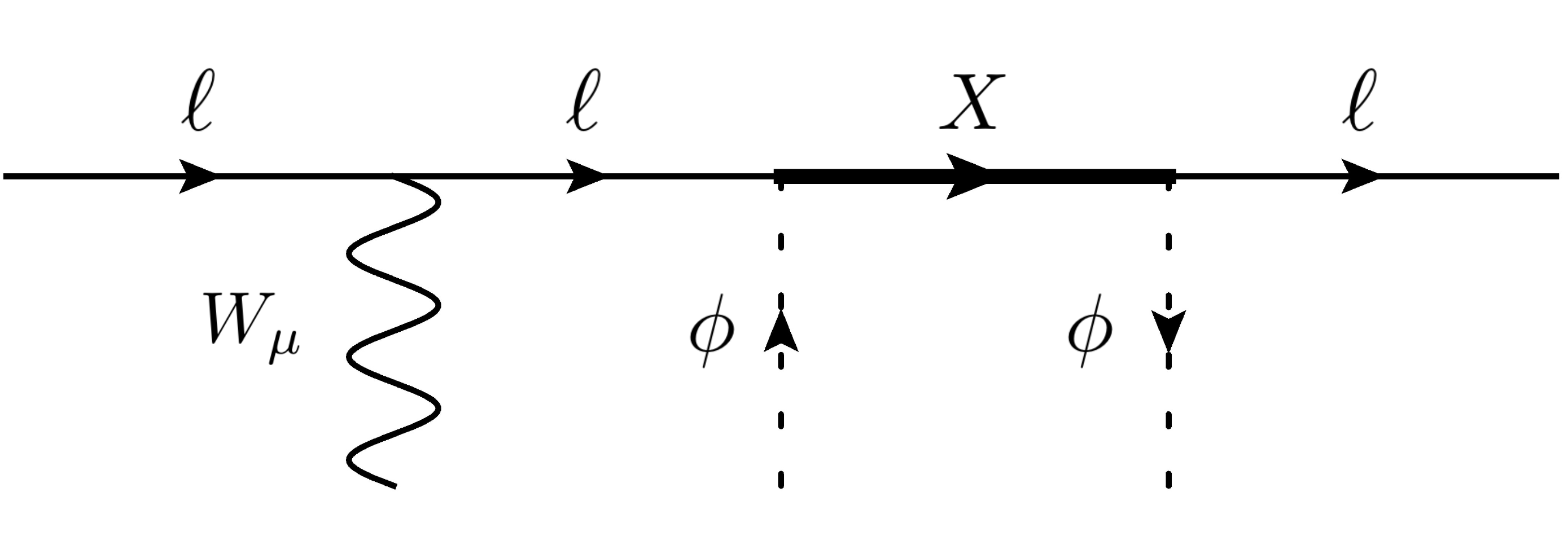} \hspace{0.9cm}
	\includegraphics[width=0.42\textwidth]{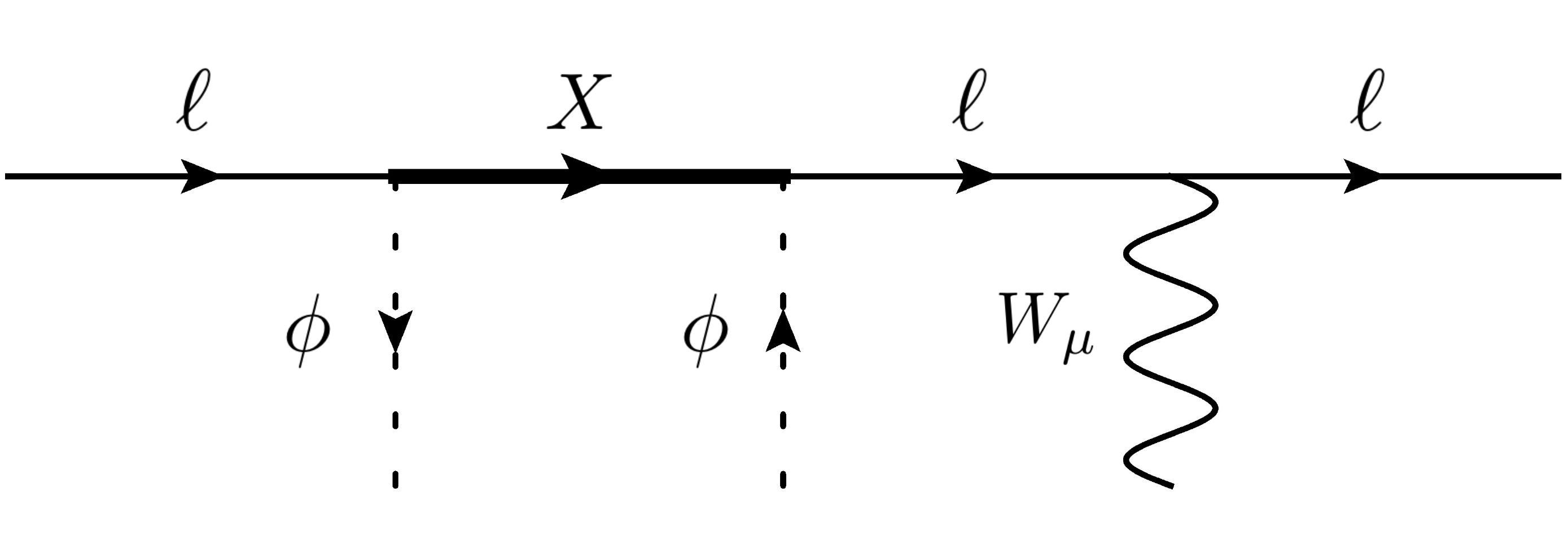} 
	\caption{Feynman diagrams giving rise to the operators $Q_{\phi \ell }^{\left( 1 \right)ij}$, $Q_{\phi \ell }^{\left( 3 \right)ij}$ and $Q_{\phi e }^{ij}$ where $X$ denotes any of the six VLLs. Note that the first diagram does not give a contribution for $N$ and $E$.  \label{FeynmanDiagrams}}
\end{figure}

Integrating out the VLLs at tree-level (see Fig.~\ref{FeynmanDiagrams}), we find the following expressions for the Wilson coefficients defined in \eq{Lagrangian}  
\begin{align}
\begin{split}
\frac{C_{\phi \ell}^{(1)ij}}{\Lambda^2} &= \frac{\lambda_N^{i}\lambda_N^{j\dagger}}{4M_N^2} -\frac{\lambda_E^{i}\lambda_E^{j\dagger}}{4M_E^2}+\frac{3}{16}\frac{\lambda_{\Sigma_0}^{i\dagger}\lambda_{\Sigma_0}^{j}}{M_{\Sigma_0}^2} -\frac{3}{16}\frac{\lambda_{\Sigma_1}^{i\dagger}\lambda_{\Sigma_1}^{j}}{M_{\Sigma_1}^2} \\
\frac{C_{\phi \ell}^{(3)ij}}{\Lambda^2} &= -\frac{\lambda_N^{i}\lambda_N^{j\dagger}}{4M_N^2} -\frac{\lambda_E^{i}\lambda_E^{j\dagger}}{4M_E^2} + \frac{1}{16}\frac{\lambda_{\Sigma_0}^{i\dagger}\lambda_{\Sigma_0}^{j}}{M_{\Sigma_0}^2} + \frac{1}{16}\frac{\lambda_{\Sigma_1}^{j\dagger}\lambda_{\Sigma_1}^{i}}{M_{\Sigma_1}^2}\\
\frac{C_{\phi \e}^{ij}}{\Lambda^2} &= \frac{\lambda_{\Delta_1}^{i\dagger}\lambda_{\Delta_1}^{j}}{2M_{\Delta_1}^2} - \frac{\lambda_{\Delta_3}^{i\dagger}\lambda_{\Delta_3}^{j}}{2M_{\Delta_3}^2}
\end{split}
\label{VLLmatch}
\end{align}
which agree with Refs.~\cite{delAguila:2008pw,deBlas:2017xtg}.

Here and in the following this notation is to be understood as
\begin{align}
\begin{aligned}
\frac{\lambda_X^i\lambda_X^{j\dagger}}{M_X^2}&=\sum_n\lambda_{X_n}^i M_{X_n}^{-2}\lambda_{X_n}^{j\dagger}\;{\rm for}\; X=N,E\,,\\
\frac{\lambda_{X}^{i\dagger}\lambda_{X}^j}{M_X^2}&=\sum_n\lambda_{X_n}^{i\dagger} M_{X_n}^{-2}\lambda_{X_n}^j\;{\rm for}\; X=\Delta_1,\,\Delta_3,\,\Sigma_0\,,\Sigma_1\,,
\end{aligned}
\end{align} 
in the case where more than one generation of VLLs is present. Without loss of generality, we assume that the mass matrices $M_X$ of the VLLs can be made real and diagonal by an appropriate choice of basis.
\smallskip

Importantly, the different representations give rise to specific patterns for the modifications of the $SU(2)_L$ gauge bosons couplings to the SM leptons. In particular, the diagonal elements even have a fixed sign:
\begin{align}
\begin{aligned}[c]
\text{N}:&\qquad C_{\phi \ell}^{(3)ii}  =\; \,-C_{\phi \ell}^{(1)ii}  < \;0,\\ 
\text{E}:&\qquad C_{\phi \ell}^{(3)ii}  = \;\;\;\;\,C_{\phi \ell}^{(1)ii}<\;0,\\
\Delta_1:&\qquad C_{\phi \ell}^{(3)ii}  =\; \;\;\;\,C_{\phi \ell}^{(1)ii}=\;0,\\
\Delta_2:&\qquad C_{\phi \ell}^{(3)ii}  = \;\;\;\;\,C_{\phi \ell}^{(1)ii}=\;0,\\
\Sigma_0:&\qquad C_{\phi \ell}^{(3)ii}  =\; \;\,\frac{1}{3}C_{\phi \ell}^{(1)ii}>\;0,\\
\Sigma_1:&\qquad C_{\phi \ell}^{(3)ii}  = -\frac{1}{3}C_{\phi \ell}^{(1)ii} >\;0,
\end{aligned}
\qquad
\begin{aligned}[c]
C_{\phi \e}^{ij}\,&=& 0,\\[3pt]
C_{\phi \e}^{ij}\,&=& 0, \\[3pt]
C_{\phi \e}^{ij}\,&>& 0,\\[3pt]
C_{\phi \e}^{ij}\,&<& 0,\\[3pt]
C_{\phi \e}^{ij}\,&=& 0,\\[3pt]
C_{\phi \e}^{ij}\,&=& 0.
\end{aligned}\label{VLCcorr}
\end{align}
The resulting modified $Z$ and $W$ couplings after EW symmetry breaking are given in Table~\ref{modSMWZcouplings} in the appendix \ref{ModWZcouplings}. Note that if the VLLs $N$ and $\Sigma_0$ are Majorana fermions, $N$ corresponds to the right-handed neutrino in the type~I seesaw~\cite{Minkowski:1977sc,Mohapatra:1979ia}, while $\Sigma_0$ corresponds to the mediator in the III mechanism~\cite{Foot:1988aq,Bajc:2006ia,Bajc:2007zf}. In this case $N$ and $\Sigma_0$ generate the neutrino mass matrices
\begin{align}
\begin{split}
	N:&\quad m_{\nu}=\frac{\lambda_{N}\lambda_{N}^T}{2M_{N}}\;v^2\,,\\
	\Sigma_0:&\quad m_{\nu}=\frac{\lambda_{\Sigma_0}^{\dagger}\lambda_{\Sigma_0}^*}{8M_{\Sigma_0}}\;v^2\,.
\end{split}
\end{align}
In general, the upper limits on the active neutrino masses set extremely stringent limits on the corresponding couplings (for a given mass of the VLLs). However, by requiring lepton number conservation~\cite{Kersten:2007vk}, as in the inverse seesaw \cite{Mohapatra:1986bd}, the effect in the neutrino masses can be avoided. In fact, it has been shown in an effective picture that such scenarios correspond to a specific pattern of the couplings $\lambda$  that allows the active neutrino masses to be small while the Dirac mass can be sizeable~\cite{Coy:2018bxr}. In the phenomenological analysis we will assume that such a mechanism is at work~\cite{Ingelman:1993ve,delAguila:2005ssc}, or simply that the VLLs $N$ and $\Sigma_0$ are Dirac fermions, meaning that the effects in modified $W$ and $Z$ couplings can be sizeable.
\medskip

\section{Observables}
\label{observables}

In this section we summarise the relevant observables for which the SM predictions are altered by the modified $W$ and $Z$ couplings, both in the EFT case and with VLLs. 

\subsection{Flavour}

Already in the EFT, modified $W$ and $Z$ couplings to leptons give rise to processes like $\ell\to \ell^\prime\gamma$ at one-loop level and can even generate $\ell\to3\ell$ and $\mu\to e$ at tree level. For the latter two, the expressions are the same in the full theory (with VLLs) and in the effective theory, while for $\ell\to \ell^\prime\gamma$ the expressions are different. 
We report the expressions for the EFT in appendix \ref{ModWZcouplings}. 
Even though all VLLs except the $N$ give rise to modified couplings of charged leptons to the $Z$ boson (see Eqs.~(\ref{Gammas}) and (\ref{VLLmatch})) and therefore contribute to $\mu\to e$ conversion, $\mu\to 3e$, $\tau\to 3\mu$, etc. already at tree-level, the latter are phase space suppressed compared to the radiative lepton decays which give competitive bounds for tau decays, even though they are induced only at the loop level.

Nonetheless, the off-diagonal elements are experimentally strongly constrained, both for the EFT~\cite{Crivellin:2013hpa,Pruna:2014asa,Crivellin:2017rmk} and the VLLs~\cite{Tommasini:1995ii,Abada:2007ux,Raidal:2008jk}.
Furthermore, since the flavour changing elements do not generate amplitudes which interfere with the SM flavour conserving observables, their effect is suppressed. Therefore, it is sufficient to consider the flavour diagonal elements $C_{\phi \ell}^{\left( 1 \right) ii}$, $C_{\phi \ell}^{\left( 3 \right) ii}$ and $C_{\phi e}^{ii}$ within the EW fit. The flavour effects that are inevitably present if there is only one generation of VLLs which couples simultaneously to at least two generations of SM leptons, will be calculated in the following. However, note that these effects can in principle be avoided by introducing multiple generations of VLLs and assuming that each SM generation mixes with at most one vector-like generation. 

\subsubsection{$\ell\to3\ell$ Processes}
In $\ell\to3\ell$ processes, we can neglect multiple flavour changes and thereby contributions to exotic decays such as $\tau^-\to e^-\mu^+e^-$ and focus on the decays involving only one flavour change. The corresponding experimental limits (at 90\% CL~\cite{Amhis:2019ckw,Bellgardt:1987du,Lees:2010ez,Hayasaka:2010np,Aaij:2014azz}) are given by
\begin{align}
\begin{split}
\operatorname{Br}(\mu \rightarrow e e e)&\leq 1.0 \times 10^{-12}\,,\\
\operatorname{Br}(\tau \rightarrow \mu \mu \mu) &\leq  1.1 \times 10^{-8}\,,\\
\operatorname{Br}(\tau \rightarrow e e e)&\leq 1.4 \times 10^{-8}\,, \\
\operatorname{Br}(\tau \rightarrow e \mu \mu) &\leq 1.6 \times 10^{-8}\,,\\
\operatorname{Br}(\tau \rightarrow \mu e e) &\leq 8.4 \times 10^{-9}\,.
\end{split}
\end{align}
The branching ratios for $\mu\to 3e$ and $\tau \to e \mu\mu$ are (here we give $\mu\to 3e$ and $\tau \to e \mu\mu$ for concreteness but the other combinations can be obtained trivially by adjusting indices)
\begin{align}
\begin{split}
{\rm Br}(\mu\to 3e) = \frac{m_{\mu}^5}{1536\pi^3m_Z^4\Gamma_{\mu}}(&2|\Gamma^{\ell L}_{e\mu}\Gamma^{\ell L}_{ee}|^2+2|\Gamma^{\ell R}_{e\mu}\Gamma^{\ell R}_{ee}|^2+|\Gamma^{\ell R}_{e\mu}\Gamma^{\ell L}_{ee}|^2+|\Gamma^{\ell L}_{e\mu}\Gamma^{\ell R}_{ee}|^2)\,,\\
{\rm Br}(\tau\to e\mu\mu) = \frac{m_{\tau}^5}{1536\pi^3m_Z^4\Gamma_{\tau}}(&|\Gamma^{\ell L}_{e\tau}\Gamma^{\ell L}_{\mu\mu}|^2+|\Gamma^{\ell R}_{e\tau}\Gamma^{\ell R}_{\mu\mu}|^2+|\Gamma^{\ell R}_{e\tau}\Gamma^{\ell L}_{\mu\mu}|^2+|\Gamma^{\ell L}_{e\tau}\Gamma^{\ell R}_{\mu\mu}|^2)\,,
\end{split}
\end{align}
with $\Gamma^{\ell L(R)}_{ij}$ given in Eq.~(\ref{Gammas}) and Eq.~(\ref{VLLmatch}) as well as in Table~\ref{modSMWZcouplings} in appendix \ref{ModWZcouplings} and $\Gamma_\mu, \, \Gamma_\tau$ are the muon and tau decay widths.

\subsubsection{Radiative Lepton Decays}
The branching ratio for $\ell_i \to \ell_f \gamma$ can be written as
\begin{align}
\label{Brmuegamma}
\Br[\ell_i \to \ell_f \gamma]&=\frac{m_{\ell_i}^3}{4\pi \, \Gamma_{i}} \big(|c_{fi}^{R} |^{2}+ |c_{if}^{R} |^{2}\big),
\end{align}
where the coefficients $c_{fi}^{R}$ are given by
\begin{align}
\begin{split}
c^{RN}_{fi}=&\frac{e }{16 \pi ^2} \,m_{\ell_i}\,\left[\lambda_N\lambda_N^\dagger\;\frac{\tilde{f}_V\left(x_N\right)-\tilde{f}_V(0)}{M_N^2}\right]_{fi}
\\
c^{RE}_{fi}=&\frac{e }{32\pi ^2 }\,m_{\ell_i}\, \left[\lambda_E\lambda_E^{\dagger }\left(\frac{\tilde{F}_\Phi\left(y_E\right)}{M_H^2}
+
\frac{-2\tilde{f}_V(0)+\tilde{F}_V\left(c_W^2 x_E\right)-2(1-2s_W^2)\,\tilde{F}_V(0)}{M_E^2}\right)\right]_{fi}
\,,\\
c^{R\Delta_1}_{fi}=&\frac{e}{32 \pi ^2}\,m_{\ell_f}\,\left[\lambda_{\Delta_1}^{\dagger }\lambda_{\Delta_1} \left(  \frac{\tilde{F}_\Phi\left(y_{\Delta_1}\right)}{M_H^2}
+\frac{\tilde{F}_V\left(c_W^2 x_{\Delta_1}\right)-4 s_W^2 \tilde{F}_V(0)}{ M_{\Delta_1}^2}\right)\right]_{fi}
\,,\\
c^{R\Delta_3}_{fi}=&\frac{e}{32 \pi ^2}\,m_{\ell_f} \,\left[\lambda_{\Delta_3}^{\dagger }\lambda_{\Delta_3}\left(\frac{ \tilde{F}_\Phi\left(y_{\Delta_3}\right) }{M_H^2}
+\frac{\tilde{F}_V\left(c_W^2 x_{\Delta_3}\right)
+4 s_W^2 \tilde{F}_V(0)}{M_{\Delta_3}^2}\right)\right]_{fi}\,,\\
c^{R\Sigma_0}_{fi}=&\frac{e}{64 \pi ^2 } \,  m_{\ell_i}\,\Bigg[\lambda_{\Sigma_0}^{\dagger }\lambda_{\Sigma_0}\Bigg(\frac{
	\tilde{F}_\Phi\left(y_{\Sigma_0}\right)}{M_H^2}
\,,\\
&\qquad\qquad\quad   
+\frac{\tilde{f}_V\left(x_{\Sigma_0}\right)+ \tilde{f}_V(0)
	+  \tilde{F}_V\left(c_W^2 x_{\Sigma_0}\right)\;
	+2\left(1-2 s_W^2\right) \tilde{F}_V(0)}{M_{\Sigma_0}^2} \Bigg)\Bigg]_{fi}\,, 
\label{mu_egamma_WCs1}
\end{split}
\end{align}
\begin{align}
\begin{split}
c^{R\Sigma_1}_{fi}=&\frac{e}{128 \pi ^2 }\, m_{\ell_i} \,\Bigg[\lambda_{\Sigma_1}^{\dagger }\lambda_{\Sigma_1}\Bigg(\frac{\tilde{F}_\Phi\left(y_{\Sigma_1}\right)}{M_H^2}
\\
   &\qquad\qquad\quad   
   + \frac{2\tilde{f}_V(0)\,+\tilde{F}_V\left(c_W^2 x_{\Sigma_1}\right)-2\left(1-2 s_W^2\right)\tilde{F}_V(0)}{M_{\Sigma_1}^2}\Bigg) \Bigg]_{fi}\,,
   \label{mu_egamma_WCs2}
   \end{split}
\end{align}
\noindent
with the loop functions being
\begin{align}
\begin{split}
\tilde f_\Phi(x)&=\frac{2x^3+3x^2-6x+1-6x^2\log x}{24(x-1)^4},\quad \tilde f_\Phi(0)=\frac{1}{24},\\
\tilde g_\Phi(x)&=\frac{x^2-1-2x\log x}{8(x-1)^3},\quad \tilde g_\Phi(0)=\frac{1}{8},\\
\tilde{F}_\Phi(x)&=\tilde f_\Phi(x)-\tilde g_\Phi(x),\quad \tilde F_\Phi(0)=-\frac{1}{12},\\
\tilde f_V(x) &= \frac{-4x^4+49x^3-78x^2 + 43x -10 - 18x^3\log x}{24(x - 1)^4},\quad \tilde f_V(0)=-\frac{5}{12},\\
\tilde g_V(x) &= \frac{-3(x^3-6x^2 + 7x -2 + 2x^2\log x)}{8(x - 1)^3},\quad \tilde g_V(0)=-\frac{3}{4}.\\
\tilde{F}_V(x)&=\tilde f_V(x)-\tilde g_V(x),\quad \tilde F_V(0)=\frac{1}{3}.
\end{split}
\end{align}
Here we show the expressions expanded up to second order in $v/M_X$ and defined
\begin{align*}
x_X\equiv \frac{M_X^2}{M_W^2},\quad y_X\equiv\frac{M_X^2}{M_H^2},\quad \quad \rm{with}\qquad X=N,\,E,\,\Delta_1,\,\Delta_3,\,\Sigma_0,\,\Sigma_1.
\end{align*}
In the presence of more than one generation of VLLs, the expressions in Eqs. (\ref{mu_egamma_WCs1})-(\ref{mu_egamma_WCs2}) are to be understood as
\begin{align}
c^{RN}_{fi}=&\frac{e }{16 \pi ^2} \,m_{\ell_i}\sum_n\,\lambda_{N_n}^f \lambda_{N_n}^{i*}\frac{\tilde{f}_V\left(x_{N_n}\right)-\tilde{f}_V(0)}{M_{N_n}^2}
\end{align}
(and similar for the others). Here $n$ runs over the number of generations of VLLs. We use the following experimental bounds on radiative leptonic decays (at 90\% C.L.) \cite{TheMEG:2016wtm}\cite{Aubert:2009ag}
\begin{align*}
\operatorname{Br}(\mu \rightarrow e \gamma)&\leq 4.2 \times 10^{-13}\,,\\
\operatorname{Br}(\tau \rightarrow e \gamma)&\leq 3.3 \times 10^{-8}\,,\\
\operatorname{Br}(\tau \rightarrow \mu \gamma)&\leq 4.4 \times 10^{-8}\,.
\end{align*}
\medskip

\subsubsection{$\mu\to e$ Conversion In Nuclei}

The induced $Ze\mu$ couplings lead to $\mu\to e$ conversion already at tree-level. These processes have stringent experimental bounds. Taking into account just this leading contribution, it is sufficient to consider the following effective Lagrangian:
\begin{align}
\mathcal{L}_{\text{eff}}=\sum_{q=u,d}\left(C_{qq}^{V\,LL}O_{qq}^{V\,LL}+C_{qq}^{V\,LR}O_{qq}^{V\,LR}\right)+L\leftrightarrow R+\hc.\,,
\end{align}
with
\begin{align}
O_{qq}^{V\,LL}&=(\bar{e}\gamma^\mu P_L\mu)(\bar{q}\gamma_\mu P_L q)\,,\qquad
O_{qq}^{V\,LR}=(\bar{e}\gamma^\mu P_L\mu)(\bar{q}\gamma_\mu P_R q)\,,
\end{align}
and 
\begin{align}
C_{qq}^{V\,LL}&=\Gamma_{e\mu}^{\ell L}\;\frac{1}{M_Z^2} \;\Gamma_{qq}^{L},\qquad\,
C_{qq}^{V\,LR}=\Gamma_{e\mu}^{\ell L}\;\frac{1}{M_Z^2} \;\Gamma_{qq}^{ R}\,,
\end{align}
where $\Gamma_{e\mu}^{\ell L/R}$ is defined in \eq{definitionZll} and given in Table~\ref{modSMWZcouplings}. The corresponding $Z$ couplings to quarks in the SM are given by
\begin{align}
\begin{aligned}
\Gamma_{uu}^L&=-\frac{g_2}{c_W}\left(\frac{1}{2}-\frac{2}{3}s_W^2\right)\,,\qquad
\Gamma_{uu}^R=\frac{2}{3}\frac{g_2 \,s_W^2}{c_W}\,,\\
\Gamma_{dd}^L&=-\frac{g_2}{ c_W}\left(-\frac{1}{2}+\frac{1}{3}s_W^2\right)\,,\quad\;\Gamma_{dd}^R=-\frac{1}{3}\frac{g_2 \,s_W^2}{c_W}\,.
\end{aligned}
\end{align}
\smallskip
Hence the transition rate $\Gamma_{\mu\to e}^N\equiv \Gamma(\mu N\to eN)$ is given by (see e.g.~\cite{Cirigliano:2009bz,Crivellin:2014cta,Crivellin:2017rmk})
\begin{align}
\Gamma_{\mu\to e}^N =4 m_\mu^5 \,\Bigg \vert \sum_{q=u,d}\left(C_{qq}^{V\;RL}+C_{qq}^{V\;RR}\right)\left(f_{Vp}^{(q)}V_N^p\,
+\, f_{Vn}^{(q)}V_N^n\right)
\Bigg\vert^2+L\leftrightarrow R\,,
\end{align}
with the nucleon vector form factors $f_{Vp}^{(u)}=2,\;f_{Vn}^{(u)}=1,\;f_{Vp}^{(d)}=1,\;f_{Vn}^{(d)}=2$ and the overlap integrals for which we use the numerical values for gold~\cite{Kitano:2002mt}
\begin{align}
V_{\text{Au}}^p=0.0974\,, \quad V_{\text{Au}}^n=0.146\,.
\end{align}
This conversion rate needs to be normalised by the capture rate~\cite{Suzuki:1987jf}
\begin{align}
\Gamma_{\text{Au}}^\text{capt}=8.7\times 10^{-18}\; \text{GeV}\,,
\end{align}
in order to be compared to the experimental 90\% C.L. limit on $\mu\to e$ conversion in gold of~\cite{Bertl:2006up}
\begin{align}
\frac{\Gamma_\text{Au}^\text{conv}}{\Gamma_\text{Au}^\text{capt}}&<7.0\times 10^{-13}\,,
\end{align}
\medskip

which makes the off-diagonal couplings $Z\mu e$ to be  negligible in our analysis.

\subsection{LFU Test}

\begin{table}[t!]
	\centering
	\begin{tabular}{l c c }
		\hline\hline
		Observable & Ref. & Measurement \\
		\hline 
		\\[-0.5cm]
		$R\left[\frac{K\rightarrow\mu\nu}{K\rightarrow e\nu}\right]\simeq|1+\frac{{v^2}}{{\Lambda^2}}C_{\phi \ell }^{\left( 3 \right)\mu\mu}-\frac{{v^2}}{{\Lambda^2}}C_{\phi \ell }^{\left( 3 \right)ee}|$ &~\cite{Pich:2013lsa} &$0.9978 \pm 0.0020$ \\[0.2 cm]		
		$R\left[\frac{\pi\rightarrow\mu\nu}{\pi\rightarrow e\nu}\right]\simeq|1+\frac{{v^2}}{{\Lambda^2}}C_{\phi \ell }^{\left( 3 \right)\mu\mu}-\frac{{v^2}}{{\Lambda^2}}C_{\phi \ell }^{\left( 3 \right)ee}|$&~\cite{Aguilar-Arevalo:2015cdf,Tanabashi:2018oca} & $1.0010 \pm 0.0009$ \\[0.2 cm]		
		$R\left[\frac{\tau\rightarrow\mu\nu\bar{\nu}}{\tau\rightarrow e\nu\bar{\nu}}\right]\simeq|1+\frac{{v^2}}{{\Lambda^2}}C_{\phi \ell }^{\left( 3 \right)\mu\mu}-\frac{{v^2}}{{\Lambda^2}}C_{\phi \ell }^{\left( 3 \right)ee}|$&~\cite{Amhis:2019ckw,Tanabashi:2018oca} & $1.0018 \pm 0.0014$ \\[0.2 cm]		
		$R\left[\frac{K\rightarrow\pi\mu\bar{\nu}}{K\rightarrow\pi e\bar{\nu}}\right]\simeq|1+\frac{{v^2}}{{\Lambda^2}}C_{\phi \ell }^{\left( 3 \right)\mu\mu}-\frac{{v^2}}{{\Lambda^2}}C_{\phi \ell }^{\left( 3 \right)ee}|$&~\cite{Pich:2013lsa} & $1.0010 \pm 0.0025$ \\[0.2 cm]		
		$R\left[\frac{W\rightarrow\mu\bar{\nu}}{W\rightarrow e\bar{\nu}}\right]\simeq|1+\frac{{v^2}}{{\Lambda^2}}C_{\phi \ell }^{\left( 3 \right)\mu\mu}-\frac{{v^2}}{{\Lambda^2}}C_{\phi \ell }^{\left( 3 \right)ee}|$&~\cite{Pich:2013lsa,Schael:2013ita} & $0.996 \pm 0.010$ \\	[0.2 cm]
		$R\left[\frac{\tau\rightarrow e\nu\bar{\nu}}{\mu\rightarrow e\bar{\nu}\nu}\right]\simeq|1+\frac{{v^2}}{{\Lambda^2}}C_{\phi \ell }^{\left( 3 \right)\tau\tau}-\frac{{v^2}}{{\Lambda^2}}C_{\phi \ell }^{\left( 3 \right)\mu\mu}|$&~\cite{Amhis:2019ckw,Tanabashi:2018oca} & $1.0010 \pm 0.0014$ \\[0.2 cm]			
		$R\left[\frac{\tau\rightarrow \pi\nu}{\pi\rightarrow \mu\bar{\nu}}\right]\simeq|1+\frac{{v^2}}{{\Lambda^2}}C_{\phi \ell }^{\left( 3 \right)\tau\tau}-\frac{{v^2}}{{\Lambda^2}}C_{\phi \ell }^{\left( 3 \right)\mu\mu}|$&~\cite{Amhis:2019ckw}& $0.9961 \pm 0.0027$ \\[0.2 cm]			
		$R\left[\frac{\tau\rightarrow K\nu}{K\rightarrow \mu\bar{\nu}}\right]\simeq|1+\frac{{v^2}}{{\Lambda^2}}C_{\phi \ell }^{\left( 3 \right)\tau\tau}-\frac{{v^2}}{{\Lambda^2}}C_{\phi \ell }^{\left( 3 \right)\mu\mu}|$&~\cite{Amhis:2019ckw} & $0.9860 \pm 0.0070$ \\[0.2 cm]
		$R\left[\frac{W\rightarrow \tau\bar{\nu}}{W\rightarrow \mu\bar{\nu}}\right]\simeq|1+\frac{{v^2}}{{\Lambda^2}}C_{\phi \ell }^{\left( 3 \right)\tau\tau}-\frac{{v^2}}{{\Lambda^2}}C_{\phi \ell }^{\left( 3 \right)\mu\mu}|$&~\cite{Pich:2013lsa,Schael:2013ita,ATLAS:2020wvq} & \begin{tabular}{@{}c @{}} $1.034 \pm 0.013|_{\text{LEP}}$\\  $0.092\pm 0.013|_{\text{ATLAS}}$ \end{tabular} \\[0.2 cm]			
		$R\left[\frac{\tau\rightarrow \mu\nu\bar{\nu}}{\mu\rightarrow e\nu\bar{\nu}}\right]\simeq|1+\frac{{v^2}}{{\Lambda^2}}C_{\phi \ell }^{\left( 3 \right)\tau\tau}-\frac{{v^2}}{{\Lambda^2}}C_{\phi \ell }^{\left( 3 \right)ee}|$&~\cite{Amhis:2019ckw,Tanabashi:2018oca} & $1.0029 \pm 0.0014$ \\[0.2 cm]			
		$R\left[\frac{W\rightarrow \tau\bar{\nu}}{W\rightarrow e\bar{\nu}}\right]\simeq|1+\frac{{v^2}}{{\Lambda^2}}C_{\phi \ell }^{\left( 3 \right)\tau\tau}-\frac{{v^2}}{{\Lambda^2}}C_{\phi \ell }^{\left( 3 \right)ee}|$&~\cite{Pich:2013lsa,Schael:2013ita} & $1.031 \pm 0.013$\\[0.2 cm]
		$R\left[\frac{B\rightarrow D^{(*)}\mu\nu}{B\rightarrow D^{(*)}e\nu}\right]\simeq|1+\frac{{v^2}}{{\Lambda^2}}C_{\phi \ell }^{\left( 3 \right)\mu\mu}-\frac{{v^2}}{{\Lambda^2}}C_{\phi \ell }^{\left( 3 \right)ee}|$&~\cite{Jung:2018lfu} & $0.989 \pm 0.012$\\[0.2 cm]
		\hline\hline
	\end{tabular}	\caption{Ratios testing LFU together with their dependence on the Wilson coefficients $C_{\phi \ell }^{\left( 3 \right)ij}$ and the corresponding experimental values. Note that here deviations from unity measures LFU violation.}\label{ObsLFU}
\end{table}

Violation of LFU in the charged current, i.e. modifications of the $W\ell\nu$ couplings, can be tested by ratios of $W$, kaon, pion and tau decays with different leptons in the final state. These ratios constrain LFU-violating effects and have reduced experimental and theoretical uncertainties. They are given by
\begin{align}
R(Y)=\frac{\mathcal{A}[Y]}{\mathcal{A}[Y]_{SM}} \,,
\end{align}
where $\mathcal{A}$ is the amplitude, and the $R(Y)$ ratio is defined in such a way that in the limit without any mixing between the SM and the VLLs, the ratios are unity. Here $Y$ labels the different observables included in our global fit which are reported in Table~\ref{ObsLFU} together with their dependence on the Wilson coefficients (see \eq{Lagrangian}) and their experimental values. Note that in all these ratios the dependence on $g_2$, the Fermi constant, etc.~drop out. In principle, the CAA could be included here via the ratio $R(V_{us})$ proposed in Ref.~\cite{Crivellin:2020lzu}. However, since we are performing a global fit, including $V_{ud}$ from beta decays and $V_{us}$ from kaon and tau decays is equivalent. Therefore, we will discuss the CAA separately later.   
\medskip

\subsection{EW Precision Observables}

\begin{table}[t!]
	\centering
	\begin{tabular}{c c}
		\hline\hline
		\begin{tabular}{c c c } 
			Observable & Ref. & Measurement  \\
			\hline
			$M_W\,[\text{GeV}]$ & ~\cite{Tanabashi:2018oca} & $80.379(12)$  \\
			$\Gamma_W\,[\text{GeV}]$ & ~\cite{Tanabashi:2018oca} & $2.085(42)$  \\
			$\text{BR}(W\to \text{had})$ & ~\cite{Tanabashi:2018oca} & $0.6741(27)$  \\
			$\text{sin}^2\theta_{\rm eff(CDF)}^{\rm e}$ & ~\cite{Aaltonen:2016nuy}  & $0.23248(52)$  \\
			$\text{sin}^2\theta_{\rm eff(D0)}^{\rm e}$ & ~\cite{Abazov:2014jti}  & $0.23146(47)$ \\
			$\text{sin}^2\theta_{\rm eff(CDF)}^{\rm \mu}$ & ~\cite{Aaltonen:2014loa}  & $0.2315(20)$ \\
			$\text{sin}^2\theta_{\rm eff(CMS)}^{\rm \mu}$ & ~\cite{Chatrchyan:2011ya}  & $0.2287(32)$ \\
			$\text{sin}^2\theta_{\rm eff(LHCb)}^{\rm \mu}$ & ~\cite{Aaij:2015lka}  & $0.2314(11)$ \\
			$P_{\tau}^{\rm pol}$ &~\cite{ALEPH:2005ab} &$0.1465(33)$ \\
			$A_{e}$ &~\cite{ALEPH:2005ab} &$0.1516(21)$  \\
			$A_{\mu}$ &~\cite{ALEPH:2005ab} &$0.142(15)$  \\
			$A_{\tau}$ &~\cite{ALEPH:2005ab} &$0.136(15)$  \\
			$\Gamma_Z\,[\text{GeV}]$ &~\cite{ALEPH:2005ab} &$2.4952(23)$ \\
		\end{tabular} &
		\begin{tabular}{c c c}
			Observable & Ref. & Measurement  \\
			\hline
			$\sigma_h^{0}\,[\text{nb}]$ &~\cite{ALEPH:2005ab} &$41.541(37)$ \\
			$R^0_{\e}$ &~\cite{ALEPH:2005ab} &$20.804(50)$ \\
			$R^0_{\mu}$ &~\cite{ALEPH:2005ab} &$20.785(33)$  \\
			$R^0_{\tau}$ &~\cite{ALEPH:2005ab} &$20.764(45)$  \\
			$A_{\rm FB}^{0, e}$&~\cite{ALEPH:2005ab} &$0.0145(25)$   \\
			$A_{\rm FB}^{0, \mu}$&~\cite{ALEPH:2005ab} &$0.0169(13)$  \\
			$A_{\rm FB}^{0, \tau}$&~\cite{ALEPH:2005ab} &$0.0188(17)$ \\
			$R_{b}^{0}$ &~\cite{ALEPH:2005ab} &$0.21629(66)$\\
			$R_{c}^{0}$ &~\cite{ALEPH:2005ab} &$0.1721(30)$ \\
			$A_{\rm FB}^{0,b}$ &~\cite{ALEPH:2005ab} &$0.0992(16)$\\ 
			$A_{\rm FB}^{0,c}$ &~\cite{ALEPH:2005ab} &$0.0707(35)$ \\
			$A_{b}$ &~\cite{ALEPH:2005ab} &$0.923(20)$ \\
			$A_{c}$ &~\cite{ALEPH:2005ab} &$0.670(27)$ \\
		\end{tabular}\\
		\hline\hline
	\end{tabular}
	\caption{EW observables included in our global fit together with their current experimental values.\label{ObsEW}}
\end{table}

The EW sector of the SM was tested with high precision at LEP \cite{Schael:2013ita,ALEPH:2005ab} and the $W$ mass has been measured with high accuracy both at Tevatron~\cite{Aaltonen:2013iut} and at the LHC~\cite{Aaboud:2017svj}. The EW sector can be completely parameterised by three Lagrangian parameters. We choose the set with the smallest experimental error: the Fermi constant ($G_F$), the fine structure constant ($\alpha$) and the mass of the $Z$ boson ($M_Z$). All other quantities and observables shown in Table~\ref{ObsEW} can be expressed in terms of these parameters and their measurements allow for consistency tests. In addition, the Higgs mass ($M_H$), the top mass ($m_t$) and the strong coupling constant ($\alpha_s$) need to be included as fit parameters, since they enter EW observables indirectly via loop effects. The theoretical predictions of Ref.~\cite{Sirlin:1980nh}, which were implemented in HEPfit~\cite{deBlas:2019okz} and are used as input parameters  in our global fit are reported in Table~\ref{ParamEW} along with their priors. \smallskip

\begin{table}[t!]
	\centering
	\begin{tabular}{l c}
		\hline\hline
		Parameter & Prior \\
		\hline
		$G_F\,\,[{\rm GeV}^{-2}]$~\cite{Tanabashi:2018oca} & $1.1663787(6) \times 10^{-5}$ \\
		$\alpha$~\cite{Tanabashi:2018oca} & $7.2973525664(17) \times 10^{-3}$ \\
		$\Delta\alpha_{\rm had}$~\cite{Tanabashi:2018oca} & $276.1(11) \times 10^{-4}$ \\
		$\alpha_s(M_Z)$~\cite{Tanabashi:2018oca} & $0.1181(11)$\\
		$M_Z\,\,[{\rm GeV}]$~\cite{ALEPH:2005ab} & $91.1875\pm0.0021$\\
		$M_H\,\,[{\rm GeV}]$~\cite{Aaboud:2018wps,CMS:2019drq} & $125.16\pm0.13$ \\
		$m_{t}\,\,[{\rm GeV}]$~\cite{TevatronElectroweakWorkingGroup:2016lid,Aaboud:2018zbu,Sirunyan:2018mlv}& $172.80\pm 0.40$ \\
		\hline\hline
	\end{tabular}
	\caption{Parameters of the EW fit together with their (Gaussian) priors. \label{ParamEW}}
\end{table}

The modifications of the $W$ and $Z$ boson couplings in \eq{Gammas} do not affect the measurements of $\alpha$ and of $M_Z$, while they do shift the value of $G_F$, which is extracted with very high precision from the decay $\mu\to e\nu\nu$. 

\noindent Taking into account that Br($\mu^+\rightarrow\e^+\nu_e\bar{\nu}_{\mu})\sim1$ we have that
\begin{align}
\frac{1}{\tau_{\mu}}=\frac{(G_F^{\mathcal{L}})^2m_{\mu}^5}{192\pi^3}(1+\Delta q)(1+C_{\phi \ell }^{\left( 3 \right)\mu\mu}+C_{\phi \ell }^{\left( 3 \right)ee})^2\,,
\end{align}
where $G_F^{\mathcal{L}}$ is the Fermi constant appearing in the Lagrangian and $\Delta q$ includes phase space, QED and hadronic radiative corrections~\cite{Fael:2020tow, Kinoshita:1958ru, vanRitbergen:1999fi, Ferroglia:1999tg}. Thus we find
\begin{align}
\begin{split}
G_F^{}&=G_F^{\mathcal{L}}(1+C_{\phi \ell }^{\left( 3 \right)\mu\mu}+C_{\phi \ell }^{\left( 3 \right)ee})\,.
\end{split}
\label{GFmod}
\end{align}
Note that within the standard set of EW observables, which is given in Table~\ref{ObsEW} and was included in our global fit, most observables are indirectly modified by \eq{GFmod} while only some of them are directly affected by the anomalous lepton-gauge boson couplings given in \eq{Gammas}.
\medskip

\subsection{Cabibbo Angle Anomaly}

As outlined in the introduction, the CAA is the disagreement between the value of $V_{ud}$ determined from beta decays and that of $V_{us}$ extracted from kaon and tau decays, once they are compared via CKM unitarity. The most precise determination of $V_{ud}$ is currently the one extracted from super-allowed $\beta$ decays~\cite{Hardy:2018zsb} and is given by
\begin{equation}
|V_{ud}|^2=\frac{2984.432(3)s}{\Ft(1+\Delta_R^V)}\,.
\end{equation}
For the $\Ft$-value we consider both the case of $\Ft=3072.07(63)s$~\cite{Hardy:2018zsb} and that of $\Ft = 3072(2)s$ including the ``new nuclear corrections'' (NNCs) that were proposed in Refs.~\cite{Seng:2018qru,Gorchtein:2018fxl}. The NCCs are included in addition to the universal electroweak corrections $\Delta_R^V$. Furthermore, there are two sets of nucleus-independent radiative corrections
\begin{align}
\Delta_R^V\big|_\text{SFGJ}&=0.02477(24)\qquad \text{\cite{Seng:2020wjq}}\,,\\
\Delta_R^V\big|_\text{CMS}&=0.02426(32)\qquad \text{\cite{Czarnecki:2019mwq}}\,.
\end{align}
Due to the smaller uncertainties in the SFGJ value, which is obtained by combining lattice QCD with dispersion relations, we will use this number in the following. Therefore, we have
\begin{align}
V_{ud}^\beta&=0.97365(15)\,, &
V_{us}^\beta &=0.2281(7)\,,\notag\\
V_{ud}^\beta\big|_\text{NNC}&=0.97366(33)\,, &
V_{us}^\beta|_{\text{NNC}} &=0.2280(14)\,,
\label{Vusbeta}
\end{align}
where we employed CKM unitarity with $|V_{ub}|=0.003683$~\cite{CKMfitter:2019,Charles:2004jd} even though the precise value of $|V_{ub}|$ is immaterial for our purpose.
\smallskip

Note that $V_{us}$ can be directly determined from the semi-leptonic kaon decays $K_{\ell 3}$. Using the compilation from Ref.~\cite{Moulson:Amherst} (updating Ref.~\cite{Moulson:2017ive}) as well as the form factor normalisation $f_+(0) = 0.9698(17)$~\cite{Carrasco:2016kpy,Bazavov:2018kjg,Moulson:Amherst}, we have that
\begin{align}
\begin{split}
V_{us}^{K_{\mu 3}}&=0.22345(54)(39)=0.22345(67)\,,\\
V_{us}^{K_{e 3}}&=0.22320(46)(39)=0.22320(61)\,,
\end{split}
\label{VusKl3}
\end{align}
where the first error refers to experiment and the second to the form factor. Here we include the determination of $V_{us}$ from the muon mode in the global fit, while the electron mode is already taken into account via the LFU ratios in Table~\ref{ObsLFU}.
\smallskip

The NP modifications to $V_{us}^{K_{\mu 3}}$ and $V_{us}^{\beta}$, including the modified couplings in \eq{Gammas} and the indirect effect of $G_F$, are
\begin{align}
\begin{split}
|V_{us}^{K_{\mu 3}}|\simeq& \; \bigg|V_{us}^{\mathcal{L}}\bigg(1-\frac{v^2}{\Lambda^2}C_{\phi \ell }^{\left( 3 \right)ee}\bigg)\bigg|\,, \\
|V_{us}^{\beta}|\simeq& \;\sqrt{1-|V_{ud}^{\mathcal{L}}|^2\bigg(1-\frac{v^2}{\Lambda^2}C_{\phi \ell }^{\left( 3 \right)\mu\mu}\bigg)^2}\,,
\end{split}
\end{align}
where $V_{us}^{\mathcal{L}}$ and $V_{ud}^{\mathcal{L}}$ are the elements of the (unitary) CKM matrix of the Lagrangian.
\smallskip

\begin{figure}[t]
	\centering
	\includegraphics[width=0.85\textwidth]{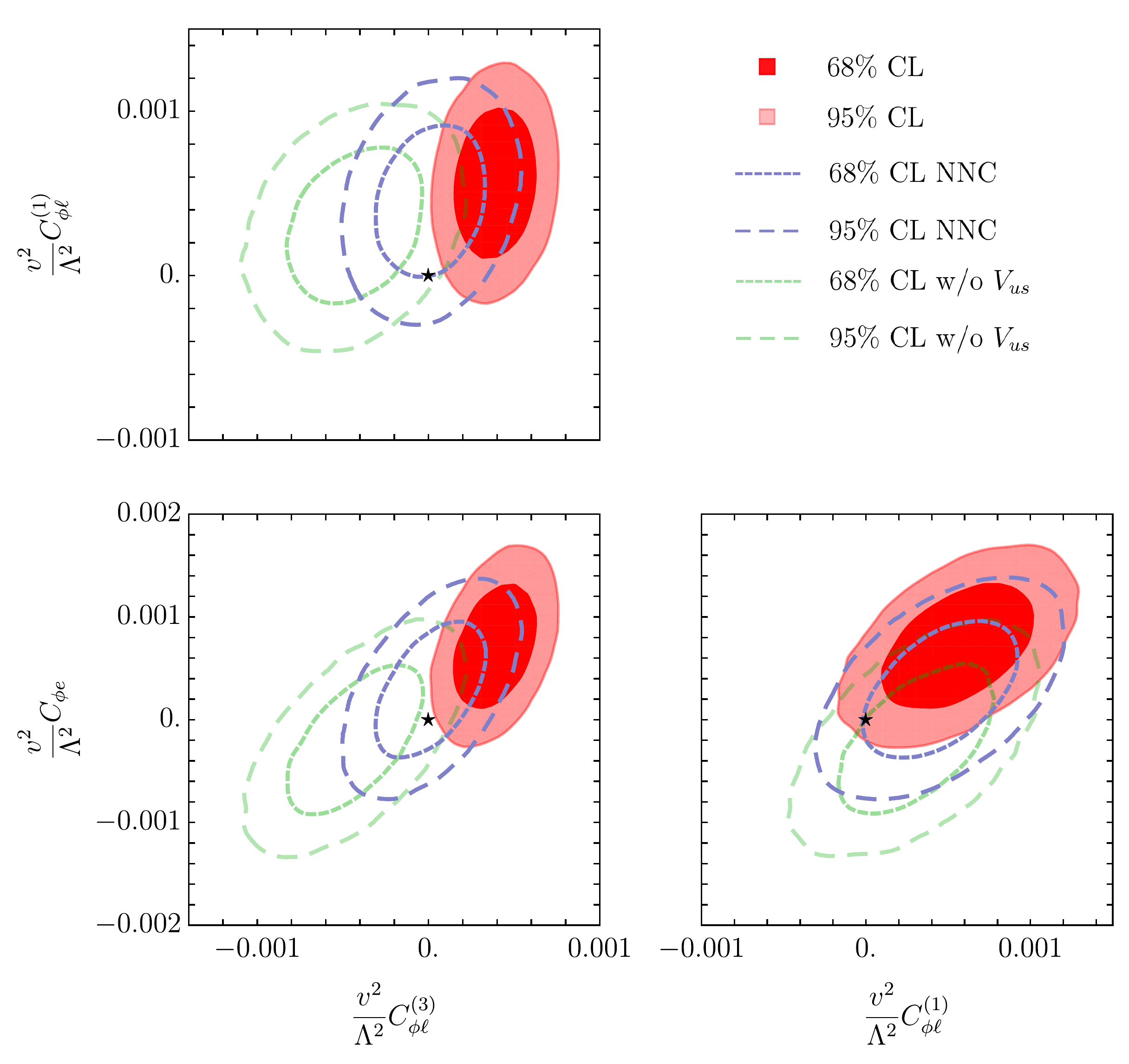} 
	\caption{Global Fit in the LFU scenario with  $C^{(3)}_{\phi\ell}\,,C^{(1)}_{\phi\ell}\,,C_{\phi e}$. The green dashed lines correspond to the standard fit (not including CKM elements), while the red regions include $V_{us}$ and $V_{ud}$ by assuming CKM unitarity. The blue dashed lines indice the region obtained if the additional NNCs are included.   \label{LFUPlot}}
\end{figure}

Regarding the purely leptonic kaon decays $K_{\ell2}$, one usually considers the ratio $K\!\to\!\mu\nu$ over $\pi\!\to\!\mu\nu$ to cancel the absolute dependence on the decay constants. This allows one to directly determine 
$V_{us}/V_{ud}$ once the ratio of decay constants $f_{K^\pm}/f_{\pi^{\pm}}$ is known and the treatment of the isospin-breaking corrections are specified~\cite{Cirigliano:2011tm,DiCarlo:2019thl}. Here, we use the recent results from lattice QCD~\cite{DiCarlo:2019thl} and at the same time adjust the FLAG average~\cite{Aoki:2019cca} back to the isospin limit 
$f_{K^\pm}/f_{\pi^{\pm}}=1.1967(18)$~\cite{Dowdall:2013rya,Carrasco:2014poa,Bazavov:2017lyh}, to obtain 
\begin{align}
V_{us}^{K_{\mu 2}}=0.22534(42)\,.
\label{VusKl2}
\end{align}
Note that this determination is insensitive to the modified $W\ell\nu$ couplings.
\smallskip

\begin{figure}[t]
	\centering
	\includegraphics[width=1.\textwidth]{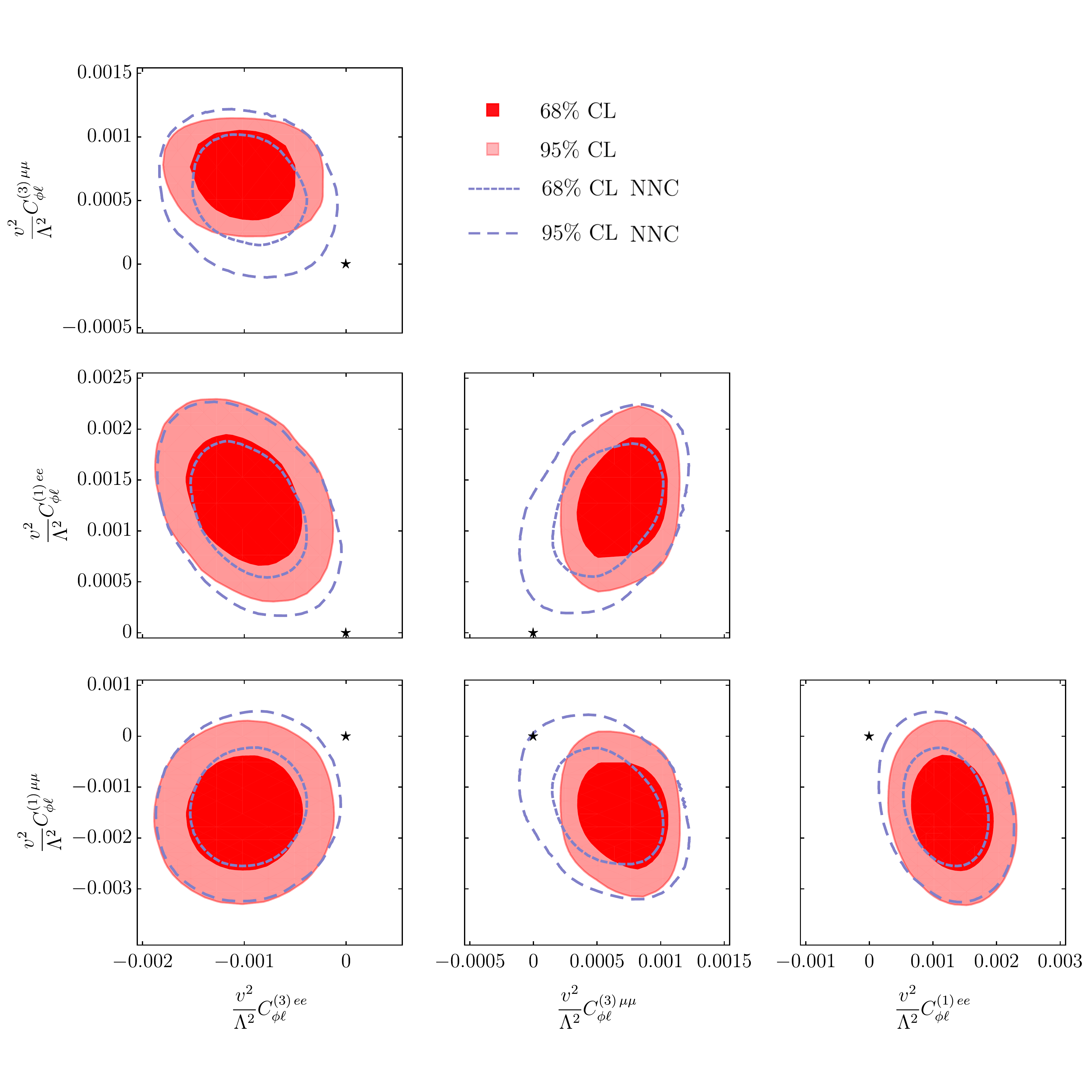} 
	\caption{Global Fit for the 6-dimensional scenario with $C^{(3)ee}_{\phi\ell}$, $C^{(3)\mu\mu}_{\phi\ell}$, $C^{(3)\tau\tau}_{\phi\ell}$, $C^{(1)ee}_{\phi\ell}$, $C^{(1)\mu\mu}_{\phi\ell}$ and $C^{(1)\tau\tau}_{\phi\ell}$ as free parameters. Here we marginalized over the Wislon coefficients with taus and do not show them explicitly since in this case there is no preference for non-zero values. The dashed lines indicate the impact of including the NNCs and the star refers to the SM point. \label{4dFit}}
\end{figure}

Alternatively, $|V_{us}|$ can be also determined from hadronic $\tau$ decays. Here the current average value for inclusive determinations is~\cite{Amhis:2019ckw} 
\begin{align}
|V_{us}^\tau| = 0.2195\pm0.0019\,.
\end{align}
Both this inclusive determination as well as the exclusive ones depend on $V_{us}/V_{ud}$, which means that there is no dependence on the modified $W$ couplings at leading order. Even though here the determination of the CKM elements is not modified by NP effects, they have an impact on the global fit as they increase the significance of the CAA. However, the exclusive modes are already included in the LFU ratios and therefore we do not include them as measurements of the CKM elements.
\medskip

\section{Analysis}\label{analysis}

Now we are in the position to perform a global analysis of all the observables discussed in the last section. We do this within a Bayesian framework using the publicly available HEPfit package~\cite{deBlas:2019okz}, whose Markov Chain Monte Carlo determination of posteriors is powered by the Bayesian Analysis Toolkit (\texttt{BAT})~\cite{Caldwell:2008fw}. With this setup we find an Information Criterion (IC)~\cite{Kass:1995} value of $\simeq 93$ for the SM.

\subsection{Model Independent Analysis}

In a first step we update the global fit assuming LFU and assess the impact of including the different determinations of $V_{us}$ on the fit. Therefore, we have only three (additional) parameters at our disposal; $C^{(3)}_{\phi\ell}\,,C^{(1)}_{\phi\ell}$ and $C_{\phi e}$. The results in all possible two-dimensional planes are given in Fig.~\ref{LFUPlot}. Interestingly, even under the assumption of LFU, including $V_{us}$ into the fit has a significant impact. In fact, without the NNCs, the 68\% C.L. regions for $C^{(3)}_{\phi\ell}$ and $C^{(1)}_{\phi\ell}$ including $V_{us}$ do not overlap with the 68\% C.L. regions for which $V_{us}$ is not included. This behaviour can be traced back to the fact that beta decays have a sensitivity to modified $W\mu\nu$ couplings, which is enhanced by $|V_{ud}|^2/|V_{us}|^2$~\cite{Crivellin:2020lzu}. Also note that while there is some preference for non-zero values of $C^{(3)}_{\phi\ell}$ and $C^{(1)}_{\phi\ell}$, $C_{\phi e}$ they are still compatible with 0 at $\simeq 2 \sigma$. Having checked explicitly that the impact of $C_{\phi e}$ on the fit is negligible, we exclude it from the following analysis which assume LFU violation.
\smallskip

\begin{figure}
	\centering
	\includegraphics[width=1\textwidth]{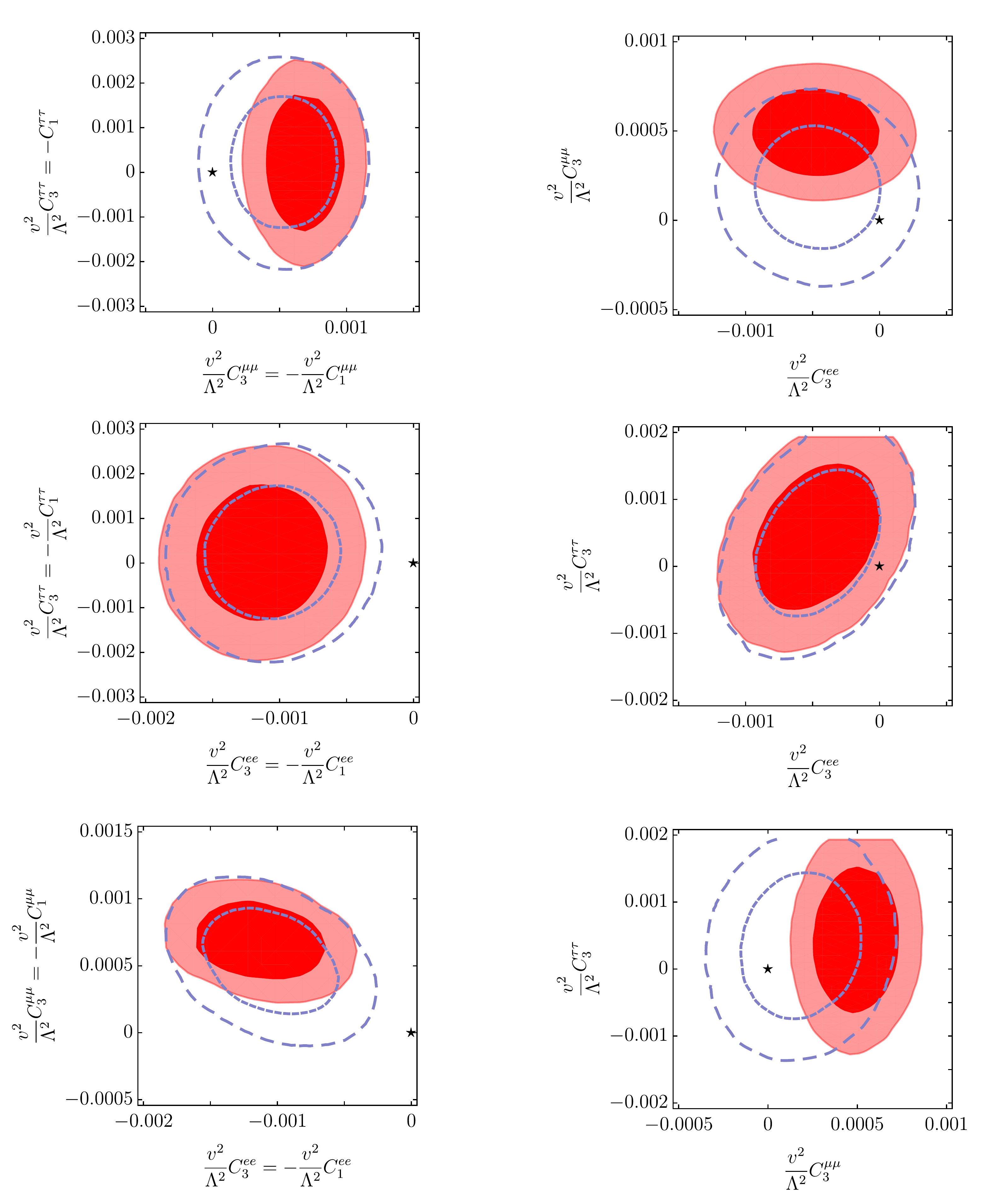}
	\caption{Global fit for the 3-dimensional scenarios $C^{(1)}_{\phi\ell}=-C^{(3)}_{\phi\ell}$ (left) and $C^{(3)}_{\phi\ell}$ only (right). Like in Fig.~\ref{4dFit}, the dashed lines indicate the effect of include the additional NNCs, the star indicates the SM point, and the regions correspond to 68\% and 95\% C.L..}
	\label{fig:C3-C1}
\end{figure}

Allowing for LFU violation, we have six free parameters in our fit, since we can neglect the flavour off-diagonal elements which are not only constrained by flavour processes but also do not lead to interference with the SM in the other observables. Furthermore, since all Wilson coefficients related to tau leptons turn out to be compatible with zero, we do not include them in Fig.~\ref{4dFit}. 
Here we again depict both the case where the NNCs are included and the case where they are neglected, finding an IC value of $83$ in both 6-dimensional scenarios, while the IC value reduces to $77$ when the tau coupling are set to zero for the outset. From these plots one can see that the pattern $C^{(1)}_{\phi\ell}=-C^{(3)}_{\phi\ell}$, already presented in Ref.~\cite{Coutinho:2019aiy}, gives a very good fit to data. This result is confirmed by an IC value of $76$ for the 3-dimensional scenario shown in Fig.~\ref{fig:C3-C1} (both for the case with NNC and without). There we also show the case of $C^{(3)}_{\phi\ell}$ only, which also provides a better fit than the SM. Here we find ${\rm IC}\simeq88$ for the scenario without NNCs and ${\rm IC}\simeq83$ with NNCs.
\medskip

\subsection{Vector Like Leptons}

\begin{figure}[t]
	\centering
	\begin{tabular}{cc}
		\includegraphics[width=1.0\textwidth]{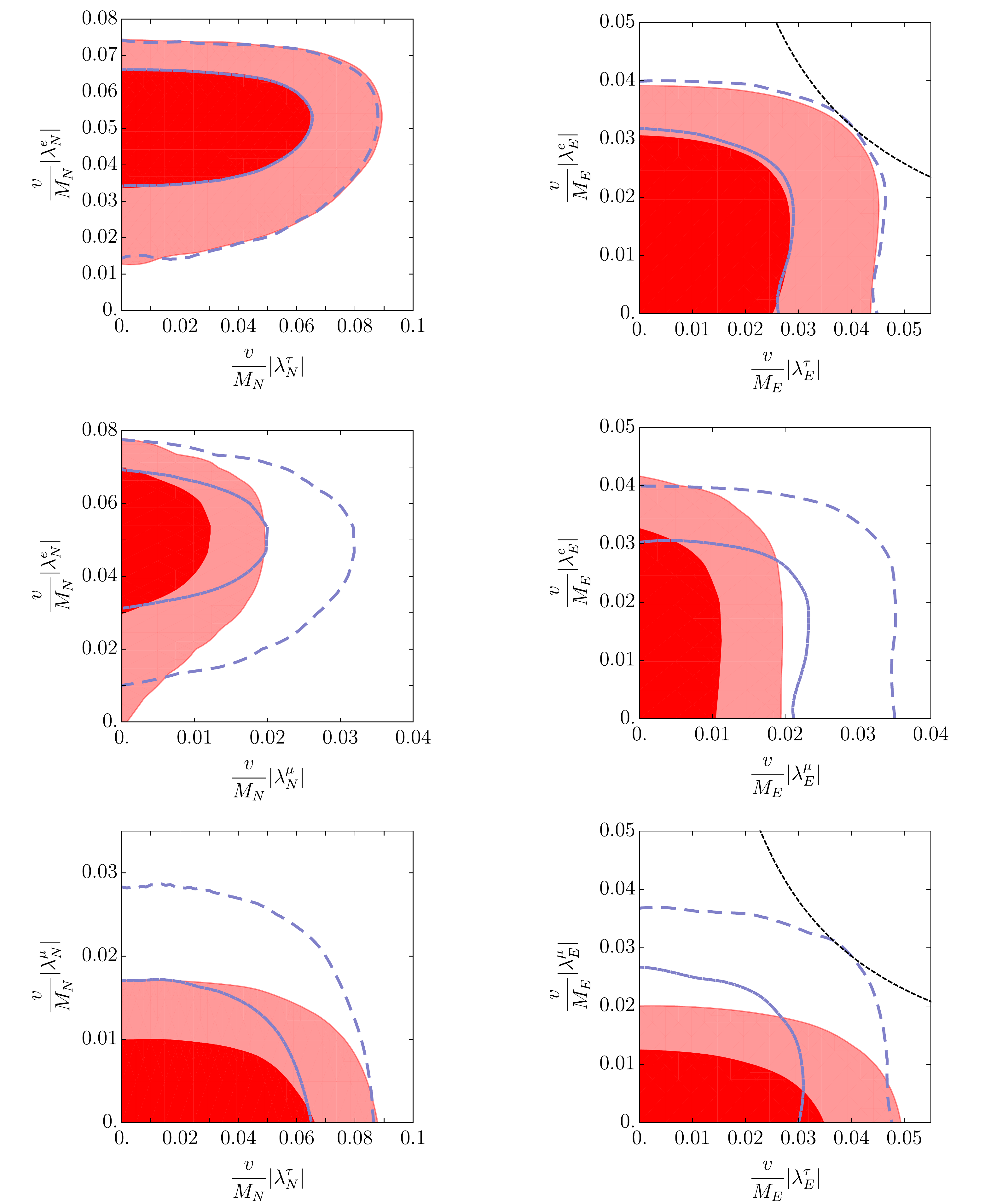} 
	\end{tabular}
	\caption{Preferred regions in parameter space for the VLLs $N$ and $E$. The color coding is the same as in Fig.~\ref{4dFit} and the black line indicates the exclusion by $\tau\to3\mu$ or $\tau\to3\mu$ in case of one generation of VLLs. The exclusions from $\mu\to e$ processes are very stringent and for better visibility shown in Fig.~\ref{Combined_mu_e_Plot}.\label{VLFit1}}
\end{figure}

\begin{figure}[t]
	\centering
	\begin{tabular}{cc}
		\includegraphics[width=1.0\textwidth]{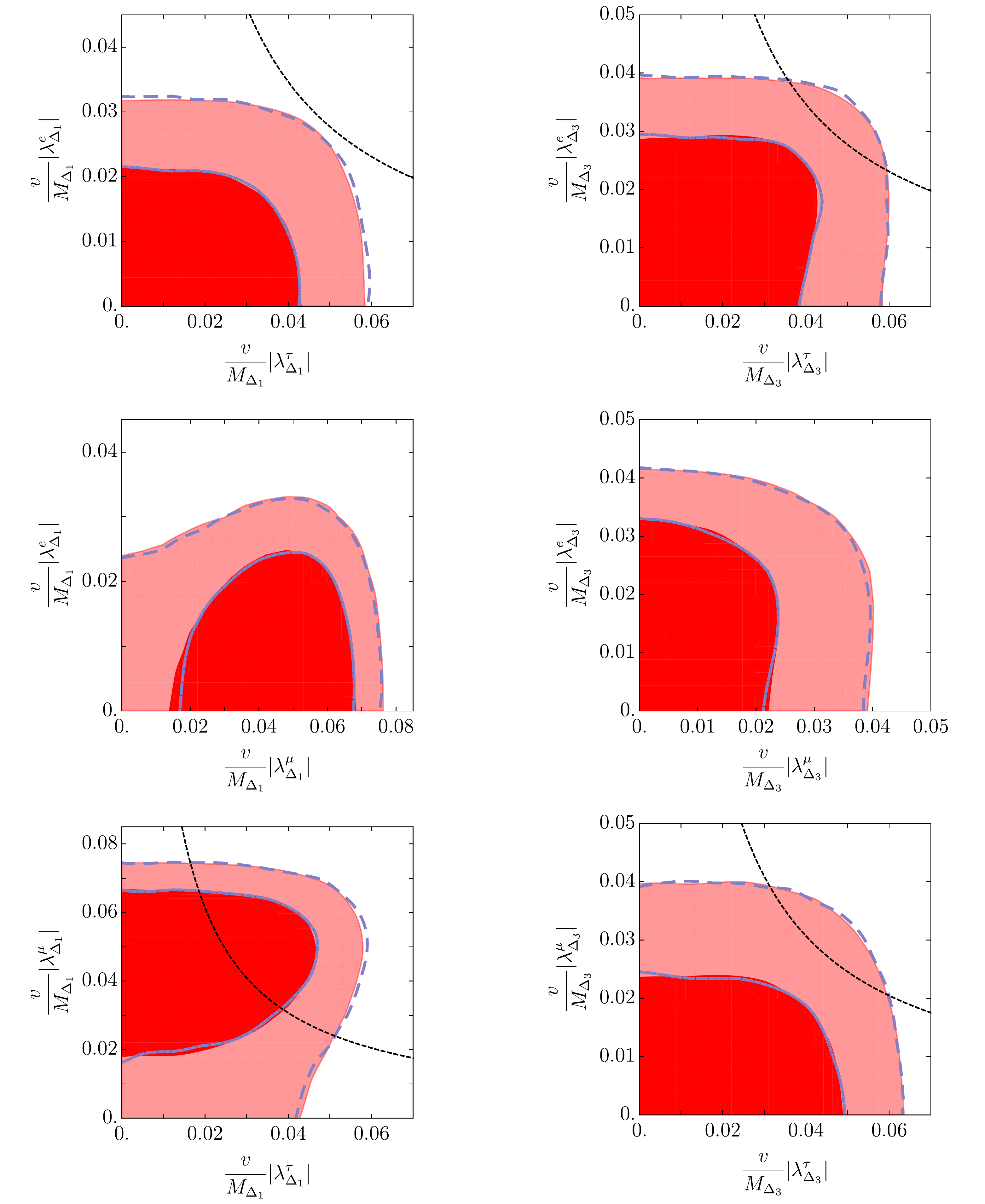}  
	\end{tabular}
	\caption{Preferred regions in parameter space for for the VLL $\Delta_3$. The color coding is the same as in Fig.~\ref{4dFit} and the black line indicates the exclusion by $\tau\to3\mu$ or $\tau\to3\mu$ in case of one generation of VLLs. The exclusions from $\mu\to e$ processes are very stringent and for better visibility shown in Fig.~\ref{Combined_mu_e_Plot}.\label{VLFit2}}
\end{figure}

\begin{figure}[t]
	\centering
	\begin{tabular}{cc}
		\includegraphics[width=1.0\textwidth]{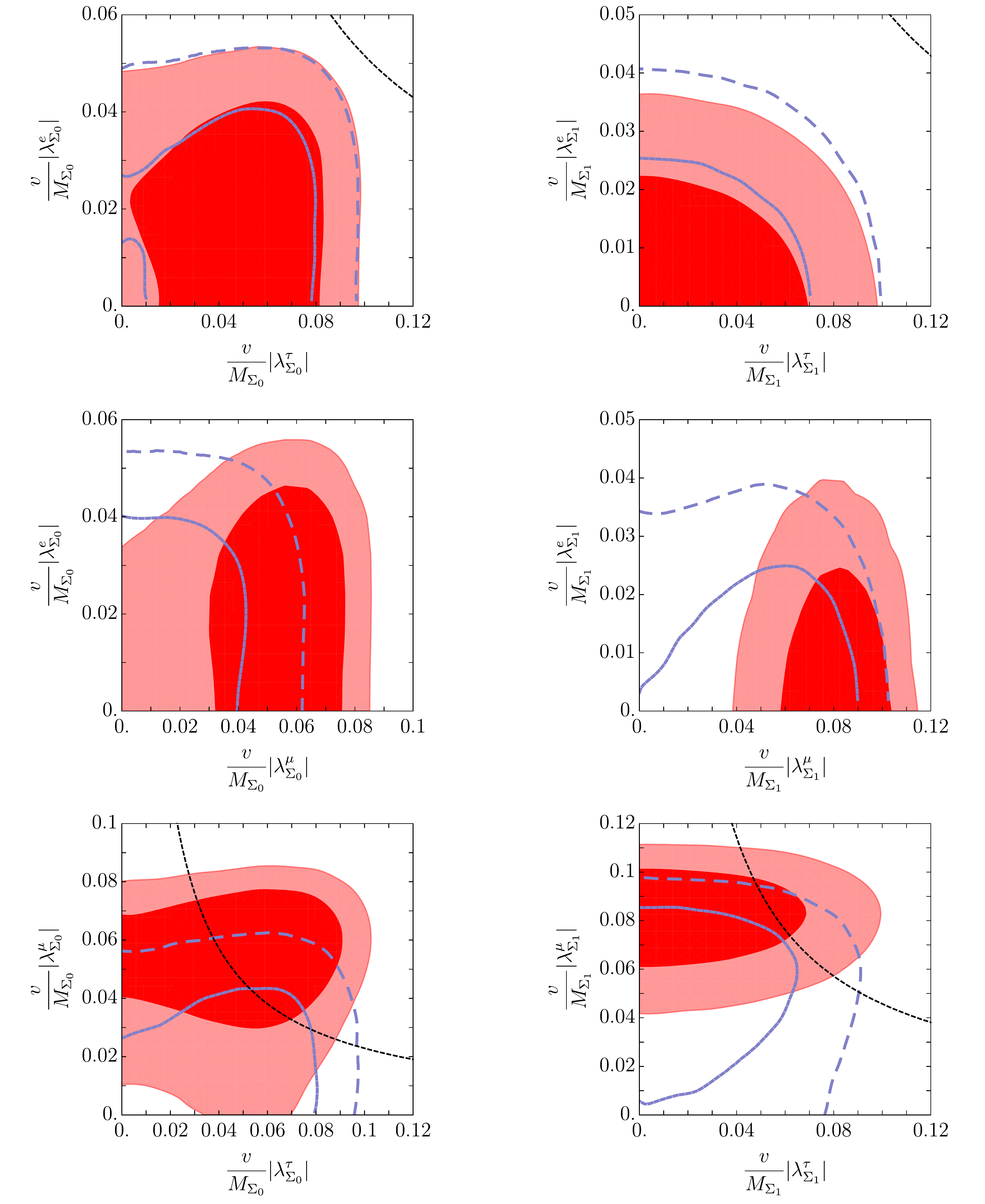} 
	\end{tabular}
	\caption{Preferred regions in parameter space for the VLLs $\Sigma_0$ and $\Sigma_1$. The color coding is the same as in Fig.~\ref{4dFit} and the black line indicates the exclusion by $\tau\to3\mu$ or $\tau\to3\mu$ in case of one generation of VLLs. The exclusions from $\mu\to e$ processes are very stringent and for better visibility shown in Fig.~\ref{Combined_mu_e_Plot}.\label{VLFit3}}
\end{figure}

\begin{figure}
	\centering
	\includegraphics[scale=0.49]{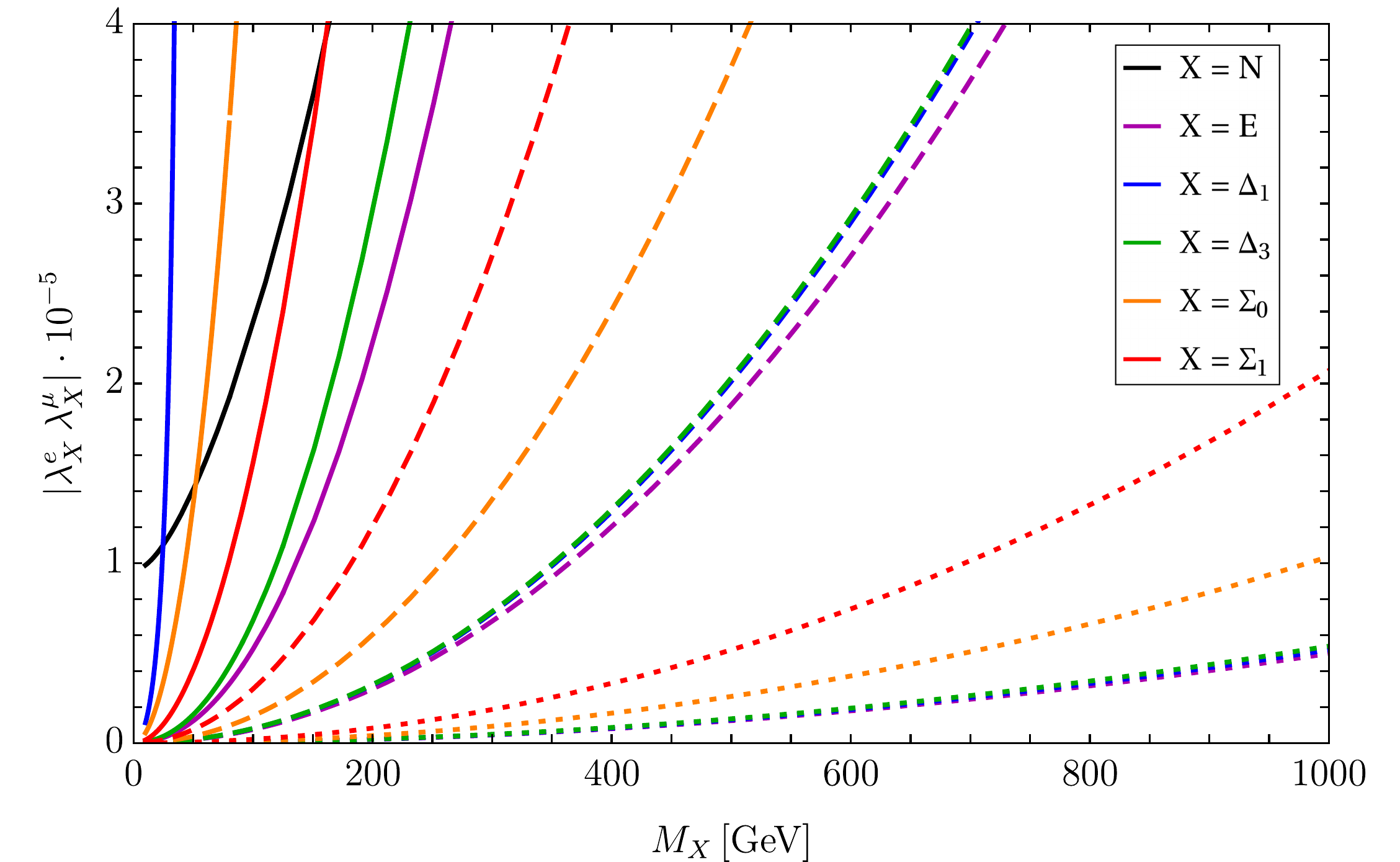}
	\caption{Upper bounds on the product $|\lambda_X^e\lambda_X^\mu|$ from the lepton flavour violating processes $\mu\to e\gamma$ (continuous lines), $\mu\to eee$ (dashed) and $\mu \to e$ conversion (dotted) as a function of the mass of the VLLs $X\,=\,N,\,E,\,\Delta_1,\,\Delta_3,\,\Sigma_0,\,\Sigma_1$. Here we assumed that only one generation of a single VLL representation is present at the same time. For $\mu\to eee$ and $\mu \to e$ conversion we only included the dominant tree-level effects induced by the modified $Z\ell\ell$ couplings. Note that therefore, $N$ only contributes to $\mu\to e\gamma$ while all other representations lead to $\mu\to3e$ and $\mu\to e$ conversion as well.
		Since for reasons of visibility only the results for the processes $\ell \to 3\ell'$ are depicted, we list the conversion factors from ${\rm Br}(\ell\to \ell ' \ell''\ell'')$ to ${\rm Br}(\ell\to 3\ell')$ (involving just one flavour off-diagonal coupling) in Table~\ref{l3l_Conversion_Factors}.}\label{Combined_mu_e_Plot}
\end{figure}

Now we turn to the patterns for the modified $W$ and $Z$ couplings to leptons obtained with VLLs. We first consider each representation separately and show the preferred regions in parameter space for each representations in Fig.~\ref{VLFit1}, Fig.~\ref{VLFit2} and Fig.~\ref{VLFit3}. Here also the bounds from $\tau\to 3\mu$ and $\tau\to 3e$ are depicted as dashed black lines. Note that the bounds from $\tau\to\mu\gamma$ and $\tau\to e\gamma$ are weaker and lie outside the displayed area. Also the flavour bounds from $\mu\to e$ processes are not shown in Fig.~\ref{VLFit1}, Fig.~\ref{VLFit2} and Fig.~\ref{VLFit3} since they are very stringent and thus would hardly be visible. Therefore, they are shown separately in Fig.~\ref{Combined_mu_e_Plot}. It is important to keep in mind that the bounds from flavour-violating processes only necessarily apply if just one generation of VLLs is present and that the bounds can be completely avoided in presence of three or more generations of VLLs. Concerning the overall goodness of the fit, note that none of the representations alone can describe data much better than the SM. This can also be seen from the obtained IC values of $93(79)$, $99(84)$,  $96(82)$, $98(84)$, $95(83)$ and $92(84)$ for N, E, $\Delta_1$, $\Delta_3$, $\Sigma_0$ and $\Sigma_1$, respectively, without (with) the NNCs.
\smallskip

\begin{figure}[t!]
	\centering
	\includegraphics[width=0.84\textwidth]{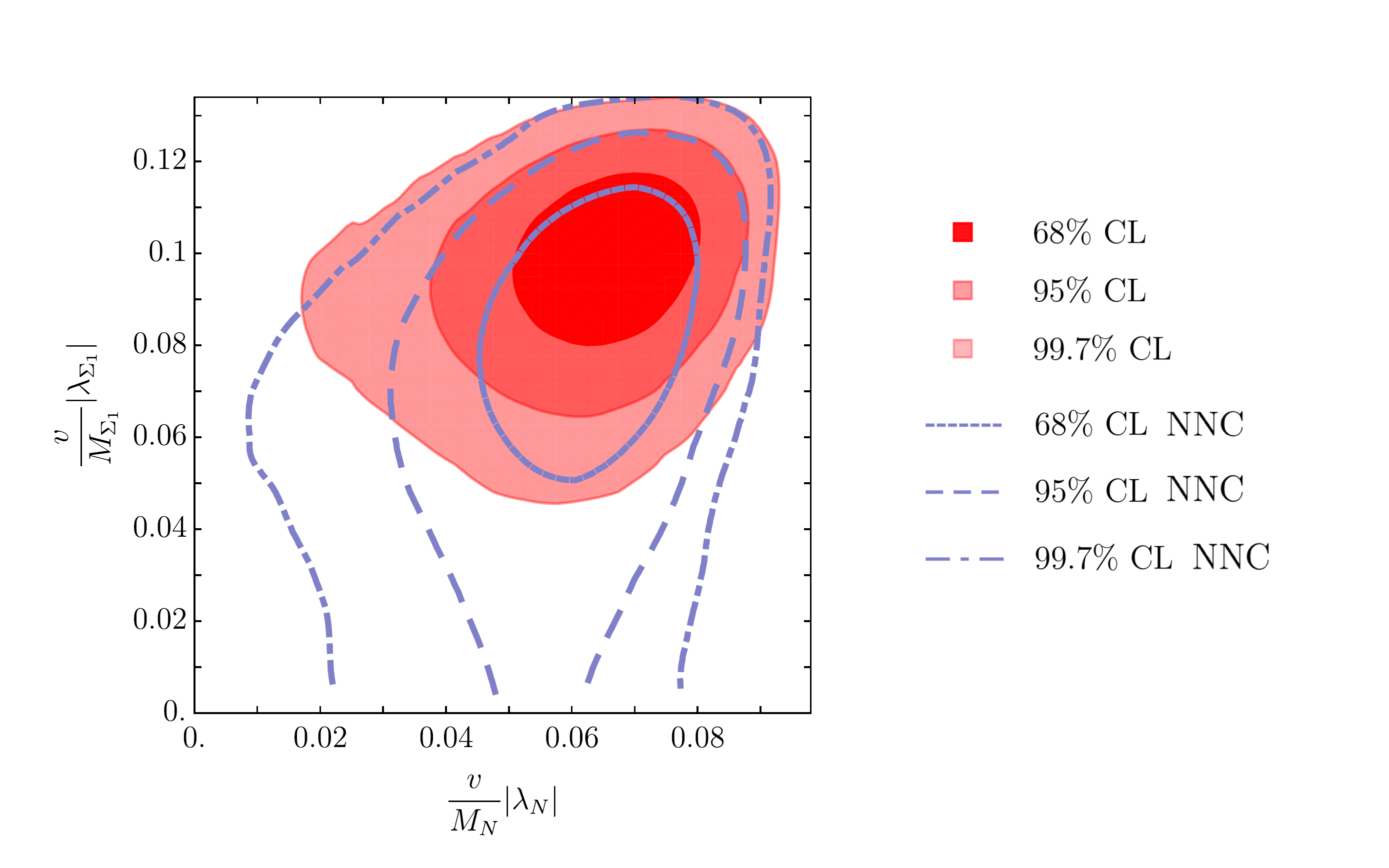} 
	\caption{Global fit in case (one generation of) the VLL $N$ couples to electrons and the VLL $\Sigma_1$ couples to muons, only. The red regions are preferred at the 68\%, 95\% and 99.7\% C.L. and the lines indicate the effect of including the NNC.\label{2DPlot}}
\end{figure}

Therefore, let us search for a simple and minimal way to combine different representations in order to obtain a good fit to data. These criteria are best met by the combination of $N$ and $\Sigma_1$, with $N$ only coupling to electrons and $\Sigma_1$ only coupling to muons. The results of the corresponding two-dimensional fit are depicted in Fig.~\ref{2DPlot}, which shows that this case is in much better agreement with data than the SM, as quantified by the IC values of 73 both in the scenario with and in the scenario without the NNCs. Since this combination of VLLs describes the data so well, we added the posteriors for the most relevant observables in Table \ref{Posteriors}.

\section{Conclusions}\label{Conclusions}

Possible modifications of the SM $Z$ and $W$ boson couplings to leptons can be most accurately constrained or determined by performing a global fit to all the available EW data. This usually includes LEP data, as well as $W$, top, and Higgs mass measurements. However, it was recently pointed out that also the CKM element $V_{us}$ (or equivalently $V_{ud}$, if CKM unitarity is employed) is affected by modified $W\ell\nu$ couplings. In fact, the interesting CAA, pointing towards a (apparent) violation of first row CKM unitarity, can be viewed as a sign of LFUV. Therefore, this anomaly does not only fall into the pattern of other hints for LFUV observed in semi-leptonic $B$ decays, but can even be explained by modified $W\ell\nu$ couplings.
\smallskip

We take this as a motivation to update the global EW fit to modified gauge boson couplings to leptons. We first study the model-independent approach where gauge-invariant dim-6 operators affect (directly) the $Z$ and $W$ couplings and find that even in the LFU case, the inclusion of CKM elements in the fit significantly impacts the results. Furthermore, for specific NP patterns like $C^{(3)}_{\phi\ell}=-C^{(1)}_{\phi\ell}$ or $C^{(3)}_{\phi\ell}$ only, the CAA leads to a preference of non-zero modifications over the SM hypothesis.
\smallskip

Moving on to the UV complete models, we studied all six representations of VLLs, which can mix, after EW symmetry breaking, with SM leptons. These different representations (under the SM gauge group) of heavy leptons lead to distinct patterns in the modifications of the $W$ and $Z$ couplings. We performed a global fit to all VLL representations separately, showing the preferred regions in parameter space which can be used to test models with VLLs against the data. In the case of a single generation of VLL, the effects on EW precision observables are correlated with the charged lepton flavour violating observables. The resulting LFV bounds (which can be avoided in presence of multiple generations of VLLs) are complementary to the regions obtained from the EW fit to $\tau-\mu$ and $\tau-e$ couplings. For $\mu-e$ couplings the bounds from flavour-violating processes are much superior to the fit-results and we show them separately in Fig.~\ref{Combined_mu_e_Plot}. Finally, while no single representation of VLL gives a fit far better than the one of the SM, we were able to identify a simple combination of VLLs which can achieve this: $N$ coupling only to electrons and $\Sigma_1$ coupling only to muons avoids the LFV constraints and agrees much better with data than the SM (by more than $4\,\sigma$). 
\smallskip

Several experimental developments are foreseen which can improve the LFU tests in Table~\ref{ObsLFU}. The J-PARC E36 experiment aims at measuring
$K\to\mu\nu/K\to e\nu$~\cite{Shimizu:2018jgs}. A similar sensitivity as in $R(V_{us})$ may be possible for $\tau\to \mu\nu\bar\nu/\tau\to  e\nu\bar\nu$  at Belle II~\cite{Kou:2018nap}, where approximately one order of magnitude more $\tau$ leptons will be produced than at BELLE or BaBar. The most promising observable is probably $\pi\to\mu\nu/\pi\to e\nu$ for which the PEN experiment anticipates an improvement by more than a factor three~\cite{Glaser:2018aat}, which would also bring the limit on $W\mu\nu$ vs $W e\nu$ modifications well below $O(10^{-3})$. Interestingly, here our $N+\Sigma_1$ model predicts a deviation from the SM expectation of more than $4\,\sigma$ which can therefore be tested in the near future.
\smallskip

\begin{table}[t!]
	\centering
	\begin{tabular}{c c c c c}
		\hline\hline
		Observable & Measurement & SM Posterior & NP Posterior & Pull\\
		\hline
		$M_W\,[\text{GeV}]$  & $80.379(12)$  & $80.363(4)$& $80.369(6)$ & \cellcolor{Gray}$0.56$\\
		\hline 
		\\[-0.5cm]
		$R\left[\frac{K\rightarrow\mu\nu}{K\rightarrow e\nu}\right]$ &$0.9978 \pm 0.0020$ & $1$& $1.00168(39)$ & $-0.80$ \\[0.2 cm]		
		$R\left[\frac{\pi\rightarrow\mu\nu}{\pi\rightarrow e\nu}\right]$& $1.0010 \pm 0.0009$& $1$ & $1.00168(39)$  &\cellcolor{Gray} $0.42$ \\[0.2 cm]		
		$R\left[\frac{\tau\rightarrow\mu\nu\bar{\nu}}{\tau\rightarrow e\nu\bar{\nu}}\right]$ & $1.0018 \pm 0.0014$& $1$ & $1.00168(39)$  &\cellcolor{Gray} $1.2$ \\[0.2 cm]					
		$|V_{us}^{K_{\mu3}}|$ &$0.22345(67)$ & $0.22573(35)$ & $0.22519(39)$  & \cellcolor{Gray}$0.77$\\[0.2 cm]
		$|V_{ud}^{\beta}|$ & $0.97365(15)$& $0.97419(8)$ & $0.97378(13)$ & \cellcolor{Gray}$2.52$\\[0.2 cm]
		\hline\hline
	\end{tabular}
	\caption{Posteriors for the observables with the largest pulls with respect to the SM in our model in which $N$ mixes with electrons and $\Sigma_1$ with muons. Note that $|V_{us}^{K_{\mu3}}|$ and $|V_{ud}^{\beta}|$ are not the Lagrangian parameters but the predictions for this CKM elements as extracted from data assuming the SM. \label{Posteriors}}
\end{table}

Clearly modified $W\ell\nu$ couplings always come together with modified $Z\ell\ell$ and/or $Z\nu\nu$ couplings. The LEP bounds on $Z\ell\ell$ couplings are already now at the per mille level~\cite{Schael:2013ita} and also the bounds on the invisible $Z$ width  (corresponding to $Z\nu\nu$ in the SM) are excellent. These bounds could be significantly improved by future $e^+e^-$ colliders such as the ILC~\cite{Baer:2013cma}, CLIC~\cite{deBlas:2018mhx}, or the FCC-ee~\cite{Abada:2019lih,Abada:2019zxq}. Furthermore, $W$ pair production will allow for a direct determination of $W\to\mu\nu/W\to e\nu$. In particular, the FCC-ee could produce up to $10^8$ $W$ bosons (compared to LEP, which produced $4\times 10^4$ $W$ bosons), leading to an increase in precision that would render a direct discovery of LFUV in $W\ell\nu$ conceivable. Furthermore, since VLLs can explain the anomalous magnetic moment of the muon (electron) and can be involved in the explanation of $b\to s\ell^+ \ell^-$ data, they are prime candidates for an extension of the SM and could also be discovered directly at the HL-LHC~\cite{ApollinariG:2017ojx} or future $e^+e^-$ colliders.

\acknowledgments
We thank Antonio Coutinho for useful discussions and help with HEPfit. This work is supported by a Professorship Grant (PP00P2\_176884) of the Swiss National Science Foundation.

\appendix
\section{Miscellaneous formulas}\label{ModWZcouplings}

The explicit expressions for the modified $W\ell\nu$, $Z\ell\ell$ and $Z\nu\nu$ couplings are given in Table~\ref{modSMWZcouplings}.
\smallskip

\noindent For the contributions of $C^{(1,3)}_{\phi\ell}$ and $C_{\phi e}$ to magnetic transitions we find
\begin{align}
\begin{split}
c^{R}_{fi}=&-\frac{e}{8 \pi ^2}\bigg[2\frac{C_{\phi \ell}^{(3)fi}}{\Lambda^2} m_{\ell_i}\tilde{f}_V(0)\\ &\qquad-\bigg(2\frac{C_{\phi e}^{fi}}{\Lambda^2} m_{\ell_f}s_W^2+\bigg(\frac{C_{\phi \ell}^{(1)fi}}{\Lambda^2} +\frac{C_{\phi \ell}^{(3)fi}}{\Lambda^2} \bigg)m_{\ell_i}(-1+2s_W^2)\bigg)(\tilde{f}_V(0)-\tilde{g}_V(0))\bigg]\,.
\end{split}
\end{align}

\noindent Since for reasons of visibility only the results for the processes $\ell \to 3\ell'$ are depicted in the 
s, we list the conversion factors from ${\rm Br}(\ell\to \ell ' 2\ell'')$ to ${\rm Br}(\ell\to 3\ell')$ (involving just one flavour off-diagonal coupling) in Table~\ref{l3l_Conversion_Factors}.
\medskip

\begin{table}[t!]
	\centering
	\begin{tabular}{|c|c|c|c|}
		\hline
		VLL & $\Gamma^{\ell \nu L}_{ij}$ & $\Gamma^{\nu}_{ij}$ \\
		\hline
		N & $-\dfrac{e}{\sqrt{2} s_W} \left(\delta_{ij}-\dfrac{v^2 \lambda_N\lambda_N^{\dagger }}{4 M_N^2}\right)$ &  $-\dfrac{e}{2\,s_W c_W}\left(\delta_{ij}-\dfrac{v^2\lambda _N^i\lambda _N^{j\dagger}}{2M_N^2}\right)$\\
		\hline
		E &$-\dfrac{e }{\sqrt{2} s_W}\left(\delta_{ij}-\dfrac{v^2\lambda_E \lambda_E^\dagger}{4M_E^2} \right)$ & -- \\
		\hline
		$\Delta_1$ & $-\dfrac{e}{\sqrt{2}s_W}\delta_{ij}$ &-- \\
		\hline
		$\Delta_3$  &  $-\dfrac{e}{\sqrt{2}s_W}\delta_{ij}$&--\\
		\hline
		$\Sigma_0$  & $-\dfrac{e}{\sqrt{2} s_W} \left(\delta_{ij}+\dfrac{v^2 \lambda_{\Sigma_0}^{i\dagger}\lambda_{\Sigma_0}^j}{16 M_{\Sigma_0}^2}\right)$&$-\dfrac{e }{2 c_W s_W}\left(\delta_{ij}-\dfrac{ v^2\lambda_{\Sigma_0}^{i\dagger} \lambda_{\Sigma_0}^j}{8 M_{\Sigma_0}^2}\right) $ \\
		\hline
		$\Sigma_1$  & $-\dfrac{e}{\sqrt{2} s_W} \left(\delta_{ij}+\dfrac{v^2 \lambda_{\Sigma_1}^{i\dagger}\lambda_{\Sigma_1}^j}{16 M_{\Sigma_1}^2}\right)$& $-\dfrac{e }{2 c_W s_W}\left(\delta_{ij}+\dfrac{v^2 \lambda_{\Sigma_1}^{i\dagger}\lambda_{\Sigma_1}^j}{4 M_{\Sigma_1}^2}\right)$\\
		\hline\hline
		VLL & $\Gamma^{\ell L}_{ij}$ & $\Gamma^{\ell R}_{ij}$\\
		\hline
		E & $ \dfrac{e}{2\,  s_W c_W } \left(\left(1-2 s_W^2\right)\delta_{ij} -\dfrac{v^2 \lambda_E^i\lambda_E^{j\dagger }}{2 M_E^2}\right)$  & $-\dfrac{e \, s_W}{c_W}\delta_{ij}$\\
		\hline
		$\Delta_1$ & $\dfrac{e }{2s_W c_W }\left(1-2 s_W^2\right) \delta_{ij}$  &$-\dfrac{e}{2 s_W c_W} \left(2 s_W^2\delta_{ij} -\dfrac{v^2 \lambda_{\Delta_1}^{i\dagger }\lambda_{\Delta_1}^j}{2
			M_{\Delta_1}^2}\right)$\\
		\hline
		$\Delta_3$ & $\dfrac{e}{2s_W c_W } \left(1-2 s_W^2\right)\delta_{ij}$ & $-\dfrac{e }{2 s_W c_W}\left(2 s_W^2 \delta_{ij}+\dfrac{v^2 \lambda_{\Delta_3}^{i\dagger }\lambda_{\Delta_3}^j}{2 M_{\Delta_3}^2}\right)$\\
		\hline
		$\Sigma_0$ & $\dfrac{e}{2 s_W c_W}\left(\left(1-2 s_W^2\right)\delta_{ij} +\dfrac{v^2 \lambda_{\Sigma_0}^{i\dagger} \lambda_{\Sigma_0}^j}{4
			M_{\Sigma_0}^2}\right)$ & $-\dfrac{e \, s_W}{c_W}\delta_{ij}$\\
		\hline
		$\Sigma_1$ & $\dfrac{e}{2 s_W c_W}\left( \left(1-2
		s_W^2\right)\delta_{ij}-\dfrac{v^2 \lambda_{\Sigma_1}^{i\dagger} \lambda_{\Sigma_1}^j}{8 M_{\Sigma_1}^2}\right)$ & $-\dfrac{e \, s_W}{c_W}\delta_{ij}$\\
		\hline
	\end{tabular}\caption{Couplings of $W$ and $Z$ to SM leptons including the corrections induced by VLLs.}\label{modSMWZcouplings}
\end{table}

\begin{table}
	\def\arraystretch{2}
	\centering
	\begin{tabular}{|c|c|c|c|c|c|}
		\hline
		VLL & N & E,$\,\Sigma_0,\,\Sigma_1$ & $\Delta_1,\,\Delta_3$ \\[5pt]
		\hline
		 $\dfrac{{\rm Br}(\ell\to \ell '\ell''\ell'')}{{\rm Br}(\ell\to 3\ell')}$ & $\;\;$--$\;\;$ & $\dfrac{8 s_W^4-4 s_W^2+1}{2 \left(6 s_W^4-4 s_W^2+1\right)}\approx 0.622 $ & $\dfrac{8 s_W^4-4 s_W^2+1}{12 s_W^4-4 s_W^2+1}\approx 0.718$\\[5pt]
		\hline
	\end{tabular}
	\caption{Conversion factors from the tree-level results for $\ell\to 3\ell'$ to the dominant (just one flavour off-diagonal vertex) tree-level contribution  to $\ell \to \ell' \ell''\ell''$. No value is given for the VLL $N$ since both ${\rm Br}(\ell\to \ell '\ell''\ell'')$ and ${\rm Br}(\ell\to 3\ell')$ are zero in this case.}\label{l3l_Conversion_Factors}
\end{table}

\newpage

\bibliographystyle{JHEP}
\bibliography{bibliography}
\end{document}